\documentclass[twocolumn,showpacs,preprintnumbers,amsmath,amssymbaps,pra,citeautoscript]{revtex4}

\usepackage{graphicx}
\usepackage{amssymb}
\usepackage{amsfonts}
\usepackage{dcolumn}
\usepackage{bm}
\usepackage{float}
\usepackage{footnote}
\usepackage{cases}
\usepackage{amsmath}
\usepackage[font=small,labelfont=bf]{caption}

\newcommand{\ket}[1]{\left| #1 \right>} 

\newcommand{\req}[1]{Eq.~(\ref{#1})}

\newcommand{\reffig}[1]{Fig. \ref{#1}}

\numberwithin{equation}{section}

\begin{document}


\title{Exact Solutions to Two-Component Many-Body Systems in One Dimension}

\author{Tianhao Ren}
 \email{tr2401@columbia.edu}
\author{Igor Aleiner}
 \email{aleiner@phys.columbia.edu}
\affiliation{Physics Department, Columbia University, New York, NY 10027, USA}

\date{\today}

\begin{abstract}
We introduce a new type of models for two-component systems in one dimension subject to exact solutions by Bethe ansatz, where the interspecies interactions are tunable via Feshbach resonant interactions. The applicability of Bethe ansatz is obtained by fine-tuning the resonant energies, and the resulting systems can be described  by introducing intraspecies repulsive and interspecies attractive couplings $c_1$ and $c_2$. This kind of systems admits two types of interesting solutions: In the regime with $c_1>c_2$, the ground state is a Fermi sea of two-strings, where the Fermi momentum $Q$ is constrained to be smaller than a certain value $Q^*$, and it provides an ideal scenario to realize BCS-BEC crossover (from weakly attractive atoms to weakly repulsive molecules) in one dimension; In the opposite regime with $c_1<c_2$, the ground state is a single bright soliton even for fermionic atoms, which reveals itself as an embedded string solution.
\end{abstract}


\maketitle

\section{Introduction}
The tremendous advances in the experimental realization of interacting quantum systems using cold atoms have recently pushed further our understanding of many-body physics beyond the simple mean-field level, which reignited the interest in low dimensional exactly solvable models. The prototype of this kind is the famous Lieb-Liniger model \cite{PhysRev.130.1605,PhysRev.130.1616}, which is now accessible to experimentalists \cite{PhysRevLett.91.250402,Kinoshita1125,PhysRevLett.93.210401,PhysRevLett.100.090402}.

As an extension of the one-component Lieb-Liniger type system, multicomponent systems host even richer many-body physics. Such systems have been realized experimentally using different hyperfine states of cold atoms, which provide us with the desired pseudospin degrees of freedom \cite{PhysRevLett.100.140401,PhysRevLett.99.190402,PhysRevA.80.023603,10.1038/nature09393}. The intra- and inter-species interactions can then be tuned via the Feshbach resonances \cite{RevModPhys.82.1225,PhysRevA.69.032705} or external potentials confining the system in one dimension \cite{PhysRevLett.91.090402,Haller1224}. On the theoretical side, the study of exactly solvable models for multicomponent systems begins with spin 1/2 fermions, which is now known as the Yang-Gaudin model \cite{GAUDIN196755,PhysRevLett.19.1312,PhysRev.168.1920}. Sutherland \cite{Sutherland_2004} and Schlottmann \cite{0953-8984-6-7-008,0953-8984-5-32-016} then made the generalization to arbitrary spins. The bosonic counterpart was also studied by various groups \cite{doi:10.1142/S0217979202011895,PhysRevA.84.033604}.

In spite of extensive studies of two-component exactly solvable models in the literature, they are all limited to the case where intra- and inter-species interactions are both repulsive or attractive. This is probably due to the non-integrability of models of simple $\delta$-contact interactions with two different coupling constants. Here we propose a new type of models for two-component systems with tunable interspecies interactions, which is subject to exact solutions by Bethe ansatz. It is of relevance not only to experimentally accessible systems such as bosonic $^{87}Rb$ quantum gases but also to fundamental theoretical problems such as BCS-BEC crossover, since, unlike the Yang-Gaudin model, it connects regimes of weakly attractive atoms to weakly repulsive molecules. Besides, it presents exotic many-body physics of which in one case the solution is a Fermi sea of two-strings. The remarkable feature is that the Fermi momentum $Q$ characterizing this sea is limited by the value $Q^*$, as the increase of the mass density at fixed interaction is accommodated by the growing density of states of two-strings. A similar phenomenon was noticed by Gurarie \cite{PhysRevA.73.033612} for a single component model with Feshbach resonance, where the system becomes unstable for small or large interactions. In the other case, an embedded string solution emerges, which means that the uniform system is unstable and it collapses into a bright soliton. This collapsing instability happens for fermionic atoms, which is contrary to the intuition that fermions won't collapse due to the Pauli exclusion principle.

The remainder of this paper is organized as follows: In Sec. \ref{sec:formalism} we present the models and discuss their integrability, then their exact solutions are worked out via the quantum inverse scattering method \cite{Samaj_2013,Korepin_1993}. Both bosonic and fermionic cases are considered and we will discover two different regimes, depending on the competition between inter- and intra-species couplings. In Sec. \ref{sec:uniform} we discuss the uniform regime with repulsion overcoming attraction, where the ground state properties and low energy excitations are derived. We also analyze the system with an external magnetic field in this regime, where a considerable portion of the phase diagram is occupied by the Fulde-Ferrel-Larkin-Ovchinnikov (FFLO) state \cite{PhysRev.135.A550,Larkin1965} and a lower critical magnetic field is found even for large densities.  In Sec. \ref{sec:string} we turn to the other regime where the ground state is a bright soliton. Finally, in the concluding section, we summarize the results and discuss possible experimental realizations and extensions.

\section{Models and Their Integrability}
\label{sec:formalism}

\subsection{Models}
Firstly, we review two famous examples of integrable models, which are relevant in our subsequent discussion of BCS-BEC crossover in one dimension. One of them, the prototype for spin 1/2 fermions, is the Yang-Gaudin model \cite{GAUDIN196755,PhysRevLett.19.1312,PhysRev.168.1920} defined by the Hamiltonian:
\begin{equation}
\label{eq:HYG}
\hat{\mathcal{H}}=\int dx\left[\partial_x\hat{\psi}^{\dagger}(x)\partial_x\hat{\psi}(x)-c_F\hat{\psi}^{\dagger}(x)\hat{\psi}^{\dagger}(x)\hat{\psi}(x)\hat{\psi}(x) \right],
\end{equation}
where $\hat{\psi}=\begin{pmatrix}\hat{\psi}_{\uparrow}\\ \hat{\psi}_{\downarrow}\end{pmatrix}$ represents the spin-1/2 fermions with mass $m_F=1/2$, and we have made the choice that $c_F>0$ corresponds to the attraction between particles. We also adopt the convention that $\hbar=1$ in this paper. It is well known that the ground state configuration of this attractive Yang-Gaudin model is a Fermi sea of singlet bound pairs, where the effective interaction between these bound pairs is still characterized by the attractive coupling $c_F$. In the limit $c_F\to 0$, the Yang-Gaudin model describes weakly bound pairs within BCS mechanism, while in the limit $c_F\to\infty$, it describes hardcore bosons instead of weakly interacting bosons.

 The second example of integrable models, the prototype for spinless bosons, is the Lieb-Liniger model \cite{PhysRev.130.1605,PhysRev.130.1616} defined by the Hamiltonian:
\begin{equation}
\label{eq:HLL}
\hat{\mathcal{H}}=\int dx\left[\partial_x\hat{\varphi}^{\dagger}(x)\partial_x\hat{\varphi}(x)+c_B\hat{\varphi}^{\dagger}(x)\hat{\varphi}^{\dagger}(x)\hat{\varphi}(x)\hat{\varphi}(x) \right],
\end{equation}
where $\hat{\varphi}$ represents the spinless bosons with mass $m_B=1/2$, and we have the opposite choice that $c_B>0$ corresponds to repulsion between particles, contrary to the Yang-Gaudin model. Now there is an interesting connection between the Yang-Gaudin model and the Lieb-Liniger model - If we identify the spinless boson as the singlet bound state of two fermions (accordingly we need to impose the mass relation such that $m_B=2m_F$), then the Yang-Gaudin model and the Lieb-Liniger model can be formally connected by just a change of sign of the coupling constant $c$. This seemingly artificial construction was proposed to be an exactly solvable model for BCS-BEC crossover in one dimension, where the connection between the two models is realized by geometric resonances \cite{PhysRevLett.93.090408,PhysRevLett.93.090405}. The Lieb-Liniger model is necessarily needed for the BEC side, because the strong coupling limit of the Yang-Gaudin model is a gas of hardcore bosons (which is also known as the fermionic super Tonks-Girardeau gas \cite{RevModPhys.85.1633}) instead of weakly interacting bosons. Although this provides a smooth crossover between the two pairing schemes, it is not satisfactory because there is no single Hamiltonian governing the behavior of the system from \req{eq:HYG} with $c_F\ll 1$ to \req{eq:HLL} with $c_B\ll 1$. Moreover, the molecule on the BEC side is unbreakable due to the quasi-1D confinement, thus the information of spin excitations is lost.

As advertised in the introduction section, in this paper we consider two-component interacting bosons and fermions with tunable interspecies interactions, which will provide an ideal scenario for one dimensional BCS-BEC crossover without the drawbacks mentioned above. Generally, the tunable interspecies interaction is realized via Feshbach resonances, where atoms are bound into molecules. Exact solutions to models with Feshbach resonances are studied in the literature for one-component interacting particles \cite{PhysRevA.73.033612,PhysRevLett.104.040402} and for noninteracting fermions in the so-called quantum three-wave interaction model \cite{doi:10.1143/JPSJ.53.1229,1742-5468-2018-2-023111}. Here we make a further step to two-component interacting systems, where the applicability of Bethe ansatz is obtained by fine-tuning the resonant energies.

We start by introducing the bosonic model, where the resonance can be viewed as a singlet bound state (molecule) of two participating bosons. It is defined by the Hamiltonian:
\begin{equation}
\label{eq:bosonicmodel}
\begin{split}
\hat{\mathcal{H}}= & \int dx \Big{\{}  \partial_x\hat{\psi}^{\dagger}\partial_x\hat{\psi}+\frac{\partial_x\hat{\Pi}^{\dagger}\partial_x\hat{\Pi}}{2m_{\Pi}}-\epsilon_0\hat{\Pi}^{\dagger}\hat{\Pi}+g\hat{\psi}^{\dagger}\hat{\psi}^{\dagger}\hat{\psi}\hat{\psi} \\
& +\left[\frac{t}{2}\left(i\partial_x\hat{\psi}^T\sigma_y\hat{\psi}\right)\hat{\Pi}^{\dagger}+h.c.\right] \Big{\}}, \\
\end{split}
\end{equation}
where $\hat{\psi}=\begin{pmatrix}\hat{\psi}_{\uparrow}\\ \hat{\psi}_{\downarrow}\end{pmatrix}$ represents the two-component bosons and $\hat{\Pi}$ represents the molecules with binding energy $\epsilon_0$. The matrix $\sigma_y$ is the $y$ component of the Pauli matrix $\bm{\sigma}=(\sigma_x,\sigma_y,\sigma_z)$. The introduction of spatial derivatives into the resonant coupling is due to the fact that the spatial part of the bosonic wave function in the singlet channel has odd parity. Also, we have adopted the convention that $m_{\psi}=1/2$, and we have left $m_{\Pi}$ unspecified. In fact, the relation between $m_{\Pi}$ and $m_{\psi}$ is dictated by Galilean invariance: Under the Galilean transformation
\begin{equation}
\begin{split}
& \partial_t \to \partial_t-v\partial_x, \\
& \hat{\psi}\to \hat{\psi}e^{-i\left( m_{\psi} vx+\frac{m_{\psi}v^2}{2}t \right)},\\
& \hat{\Pi}\to \hat{\Pi}e^{-i\left( m_{\Pi} vx+\frac{m_{\Pi}v^2}{2}t \right)},
\end{split}
\end{equation}
the action $S=\int dxdt ~\left(i\hat{\psi}^{\dagger}\partial_t\hat{\psi}+i\hat{\Pi}^{\dagger}\partial_t\hat{\Pi}\right)-\int dt~\hat{\mathcal{H}}$ of the system will remain unchanged apart from a constant shift, provided we impose the relation that $m_{\Pi}=2m_{\psi}$. Thus we can simply substitute $m_{\Pi}=1$ into \req{eq:bosonicmodel}. There is one more point that we need to pay attention to, which is the conservation of particle number. This means that the operator $\hat{N}$ defined below commutes with the Hamiltonian $\hat{\mathcal{H}}$:
\begin{equation}
\label{eq:Number1}
\hat{N}=\int dx~\left( \hat{\psi}^{\dagger}\hat{\psi}+2\hat{\Pi}^{\dagger}\hat{\Pi} \right), ~~~ [\hat{N},\hat{\mathcal{H}}]=0.
\end{equation}
The definition of $\hat{N}$ takes into account the resonant coupling processes where two bosons with opposite pseudospin transform into one molecule or vice versa. This commutability can be achieved by imposing the following commutation relations:
\begin{equation}
\begin{split}
& [\hat{\psi}_{\sigma}(x),\hat{\psi}^{\dagger}_{\sigma'}(x')]=\delta_{\sigma\sigma'}\delta(x-x'), ~~~ \sigma,\sigma'=\uparrow,\downarrow\\
& [\hat{\Pi}(x),\hat{\Pi}^{\dagger}(x')]=\delta(x-x'),
\end{split}
\end{equation}
and all the other commutators give out zero.

The fermionic model can be constructed similarly, with the introduction of both scaler and vector resonances. It is defined by the Hamiltonian:
\begin{equation}
\label{eq:fermionicmodel}
\begin{split}
\hat{\mathcal{H}}= & \int dx \Big{\{}  \partial_x\hat{\psi}^{\dagger}\partial_x \hat{\psi}+\frac{1}{2}\partial_x\hat{\bm{\Xi}}^{\dagger}\cdot\partial_x\hat{\bm{\Xi}} +\frac{1}{2}\partial_x\hat{\Pi}^{\dagger}\partial_x\hat{\Pi} \\
& -\epsilon_{\Xi}\hat{\bm{\Xi}}^{\dagger}\cdot\hat{\bm{\Xi}}-\epsilon_{\Pi}\hat{\Pi}^{\dagger}\hat{\Pi}+g\hat{\psi}^{\dagger}\hat{\psi}^{\dagger}\hat{\psi}\hat{\psi}\\
&+\left[\frac{t_{\Xi}}{2}\left(i\partial_x\hat{\psi}^T\bm{\sigma}\sigma_y\hat{\psi}\right)\cdot\hat{\bm{\Xi}}^{\dagger}+h.c.\right]\\
&+\left[\frac{t_{\Pi}}{2}\left(i\hat{\psi}^T\sigma_y\hat{\psi}\right)\cdot\hat{\Pi}^{\dagger}+h.c.\right] \Big{\}},
\end{split}
\end{equation}
where $\hat{\psi}=\begin{pmatrix}\hat{\psi}_{\uparrow}\\ \hat{\psi}_{\downarrow}\end{pmatrix}$ represents the spin-1/2 fermions, $\hat{\bm{\Xi}}$ represents the vector resonances with binding energy $\epsilon_{\Xi}$ and $\hat{\Pi}$ represents the scalar resonances with binding energy $\epsilon_{\Pi}$. Also we adopt the convention that $m_{\psi}=1/2$, then $m_{\Xi}=m_{\Pi}=1$ is required by the Galilean invariance, just as what we have discussed previously. Again we need to be careful about the conservation of particle number:
\begin{equation}
\label{eq:Number2}
\hat{N}=\int dx~\left( \psi^{\dagger}\psi+2\hat{\bm{\Xi}}^{\dagger}\cdot\hat{\bm{\Xi}}+2\hat{\Pi}^{\dagger}\hat{\Pi} \right), ~~~[\hat{N},\hat{\mathcal{H}}]=0.
\end{equation}
This can be achieved by imposing the following commutation and anticommutation relations:
\begin{equation}
\begin{split}
& \{\hat{\psi}_{\sigma}(x),\hat{\psi}^{\dagger}_{\sigma'}(x')\}=\delta_{\sigma\sigma'}\delta(x-x'),\\
& \{\hat{\psi}^{\dagger}_{\sigma}(x),\hat{\psi}^{\dagger}_{\sigma'}(x')\}=0, ~~~ \{\hat{\psi}_{\sigma}(x),\hat{\psi}_{\sigma'}(x')\}=0\\
& [\hat{\Xi}_{\mu}(x),\hat{\Xi}^{\dagger}_{\mu'}(x')]=\delta_{\mu\mu'}\delta(x-x'), \\
& [\hat{\Pi}(x),\hat{\Pi}^{\dagger}(x')]=\delta(x-x'),
\end{split}
\end{equation}
and all the other commutators give out zero. In the above equations, the spin labels $\sigma,\sigma'$ take values of $\uparrow,\downarrow$, and the polarization labels $\mu,\mu'$ take values of $+,-,z$. The relation between polarization labels and vector labels is as follows:
\begin{equation}
\hat{\Xi}_{+}=\frac{1}{\sqrt{2}}(\hat{\Xi}_x+i\hat{\Xi}_y), ~~~ \hat{\Xi}_{-}=\frac{1}{\sqrt{2}}(\hat{\Xi}_x-i\hat{\Xi}_y).
\end{equation}
The bosonic and fermionic models introduced here can be solved via Bethe ansatz if we fine tune the resonant energies $\epsilon_0,\epsilon_{\Xi},\epsilon_{\Pi}$, and both of them can be effectively described as a two-component system with intraspecies repulsion and interspecies attraction. By tuning the strength of attraction from vanishingly small toward the strength of repulsion, the system first shows BCS-type pairing behavior, then it develops toward the fermionic super Tonks-Girardeau gas regime, and finally, it turns into the weakly interacting bosons regime and shows BEC-type pairing behavior. Thus we can have a single Hamiltonian governing the whole range of BCS-BEC crossover, and there is no geometric confinement preventing the breaking of bound pairs. What is more, if we tune the strength of attraction beyond repulsion, we will enter into a new regime where the uniform configuration is unstable and the system collapses into a bright soliton. Now let us discuss these interesting physics one by one, starting from the integrability.

\subsection{Integrability}
Integrable models with internal degrees of freedom can be solved using the quantum inverse scattering method \cite{Samaj_2013,Korepin_1993}. The essential point is to construct the two-body $S$-matrix which fulfills the Yang-Baxter equation. For a time inversion and space inversion invariant and species-conserving model in free space, the two-body $S$-matrix in pseudospin subspace $\{\uparrow, \downarrow\}$ assumes the following general form:
\begin{equation}
\label{eq:Smatrix}
S(k)=\begin{pmatrix}
a(k) & 0 & 0 & 0\\
0 & b(k) & c(k) & 0\\
0 & c(k) & b(k) & 0\\
0 & 0 & 0 & a(k)
\end{pmatrix},
\end{equation}
where the relative momentum $k=k_2-k_1$ is the difference in momentum between the two scattering particles. The requirement of unitarity
\begin{equation}
S^{\dagger}(k)S(-k)=S(k)S^{\dagger}(-k)=I
\end{equation}
together with the Yang-Baxter equation
\begin{equation}
\begin{split}
& \sum_{\sigma_1'\sigma_2'\sigma_3'}S^{\sigma_1\sigma_2}_{\sigma_1'\sigma_2'}(k_1-k_2)S^{\sigma_1'\sigma_3}_{\sigma_1''\sigma_3'}(k_1)S^{\sigma_2'\sigma_3'}_{\sigma_2''\sigma_3''}(k_2)\\
=& \sum_{\sigma_1'\sigma_2'\sigma_3'}S^{\sigma_2\sigma_3}_{\sigma_2'\sigma_3'}(k_2)S^{\sigma_1\sigma_3'}_{\sigma_1'\sigma_3''}(k_1)S^{\sigma_1'\sigma_2'}_{\sigma_1''\sigma_2''}(k_1-k_2)
\end{split}
\end{equation}
then put severe constraints on the $S$-matrix elements:
\begin{equation}
\label{eq:YangBaxter}
\begin{split}
	& a(k)a(-k)=1, a(k)=b(k)+c(k),\frac{a(k)}{b(k)}+\frac{a(-k)}{b(-k)}=2,\\
	& \frac{a(k_1)b(k_2)-a(k_2)b(k_1)}{b(k_2-k_1)}=\frac{c(k_1)c(k_2)}{c(k_2-k_1)}.
\end{split}
\end{equation}
As a result, the integrability can be checked by identifying the $S$-matrix elements in \req{eq:Smatrix} and checking the validity of \req{eq:YangBaxter}. Then the exact solutions can be explicitly constructed using the algebraic Bethe ansatz \cite{Samaj_2013,Korepin_1993}, and the resulting Bethe ansatz equations for an eigenstate with $M$ out of $N$ particles placed spin-down are as follows:
\begin{equation}
\label{eq:Nested}
\begin{split}
& \prod_{n=1}^N\frac{a(\Lambda_{\alpha}-k_n)}{b(\Lambda_{\alpha}-k_n)}=\prod_{\substack{\beta=1\\\beta\neq \alpha}}^M\frac{a(\Lambda_{\alpha}-\Lambda_{\beta})b(\Lambda_{\beta}-\Lambda_{\alpha})}{a(\Lambda_{\beta}-\Lambda_{\alpha})b(\Lambda_{\alpha}-\Lambda_{\beta})},\\
& (\mp)^{N-1}\exp(-ik_jL)=\prod_{n=1}^Na(k_j-k_n)\prod_{\alpha=1}^M\frac{a(\Lambda_{\alpha}-k_j)}{b(\Lambda_{\alpha}-k_j)},
\end{split}
\end{equation}
where $L$ is the size of the system, the charge rapidities $k_n, n=1,2,\cdots,N$ are the physical momenta, and the spin rapidities $\Lambda_{\alpha},\alpha=1,2,\cdots,M$ are auxiliary parameters (thus the ansatz is also called nested Bethe ansatz). Also in the second set of equations, the upper sign is for bosons, and the lower sign is for fermions.

Let us first remind the reader about the Bethe ansatz equations for the Yang-Gaudin model and the Lieb-Liniger model and then turn to the present models defined in Eqs. (\ref{eq:bosonicmodel}) and (\ref{eq:fermionicmodel}).

The $S$-matrix elements for the Yang-Gaudin model described in \req{eq:HYG} are \cite{Samaj_2013,Korepin_1993}:
\begin{equation}
\label{eq:YG}
a(k)=1, ~~~ b(k)=\frac{k}{k-ic_F}, ~~~ c(k)=\frac{-ic_F}{k-ic_F}.
\end{equation}
This clearly fulfills the integrability criterion as specified in \req{eq:YangBaxter}. Then it is straightforward to substitute \req{eq:YG} into \req{eq:Nested} to obtain the Bethe ansatz equations:
\begin{equation}
\begin{split}
& \prod_{j=1}^{N}\left( \frac{\Lambda_{\alpha}-k_j-ic_F'}{\Lambda_{\alpha}-k_j+ic_F'} \right)=-\prod_{\beta=1}^M\left( \frac{\Lambda_{\alpha}-\Lambda_{\beta}-ic_F}{\Lambda_{\alpha}-\Lambda_{\beta}+ic_F} \right), \\
& \exp(ik_jL)=\prod_{\alpha=1}^M\left( \frac{k_j-\Lambda_{\alpha}-ic_F'}{k_j-\Lambda_{\alpha}+ic_F'} \right),
\end{split}
\end{equation}
where $c_F'=c_F/2$ and we have made the conventional shift $\Lambda_{\alpha}\to \Lambda_{\alpha}+ic_F'$. The ground state configuration of this attractive Yang-Gaudin model (with even number of particles) is known to be a Fermi sea of two-string solutions with the structure:
\begin{equation}
k_{\alpha,1}=\Lambda_{\alpha}+ic'_F, ~~~ k_{\alpha,2}=\Lambda_{\alpha}-ic'_F,
\end{equation}
where $\alpha=1,2,\cdots,M$ with $M=N/2$, and the Bethe ansatz equations reduce into equations for the center momenta $\Lambda_{\alpha}$:
\begin{equation}
\exp(2i\Lambda_{\alpha}L)=(-)\prod_{\beta=1}^M\left( \frac{\Lambda_{\alpha}-\Lambda_{\beta}-ic_F}{\Lambda_{\alpha}-\Lambda_{\beta}+ic_F} \right).
\end{equation}
Because the center momenta $\Lambda_{\alpha}$ in the ground state configuration are all real, we can take the logarithm of the above equations:
\begin{equation}
\label{eq:DiscreteBethe}
2\Lambda_{\alpha}L=2\pi J_{\alpha}-\sum_{\beta=1}^M\theta(\Lambda_{\alpha}-\Lambda_{\beta}),
\end{equation}
where $J_{\alpha}$s are consecutive integers or half-odd integers depending on:
\begin{equation}
J_{\alpha}=\frac{M+1-2\alpha}{2}, ~~~ \alpha=1,2,\cdots,M.
\end{equation}
The phase-shift function $\theta(\Lambda-\Lambda')$ is defined as:
\begin{equation}
\label{eq:PhaseYG}
\theta(\Lambda-\Lambda')=-2\arctan\left(\frac{\Lambda-\Lambda'}{c_F}\right).
\end{equation}
In the thermodynamic limit where $N\to\infty,L\to\infty$ with fixed density $n_F=N/L$, \req{eq:DiscreteBethe} can be replaced by an integral equation for the density of state $\sigma(\Lambda_j)=1/L(\Lambda_{j+1}-\Lambda_j)$, following Hult\'{e}n's continualization procedure \cite{Hulten}:
\begin{equation}
\label{eq:IntYG}
\sigma(\Lambda)+\frac{c_F}{\pi}\int_{-Q}^Qd\Lambda'\frac{\sigma(\Lambda')}{(\Lambda-\Lambda')^2+c_F^2}=\frac{1}{\pi},
\end{equation}
where $Q$ is the Fermi momentum of the ground state, and it is determined by the density $n_F$ via the relation:
\begin{equation}
\frac{n_F}{2}=\int_{-Q}^Qd\Lambda~\sigma(\Lambda).
\end{equation}
After solving the above integral equation, the physical quantities of the ground state can then be calculated in terms of the density of states $\sigma(\Lambda)$.

The Lieb-Liniger model as described in \req{eq:HLL} in fact has simpler structure compared with the Yang-Gaudin model, because it is a single-component model. The resulting ground state configuration with $N$ particles fulfills the Bethe ansatz equations:
\begin{equation}
k_jL=2\pi I_j-\sum_{n=1}^N\theta(k_j-k_n),
\end{equation}
where $I_j$s are consecutive integers or half-odd integers depending on:
\begin{equation}
I_j=\frac{N+1-2j}{2}, j=1,2,\cdots,N.
\end{equation}
The corresponding phase-shift function is defined as
\begin{equation}
\label{eq:PhaseLL}
\theta(k-k')=2\arctan\left( \frac{k-k'}{c_B}\right),
\end{equation}
and the integral equation for the density of states $\rho(k)=1/L(k_{j+1}-k_j)$ in the thermodynamic limit is:
\begin{equation}
\label{eq:IntLL}
\rho(k)-\frac{c_B}{\pi}\int_{-q}^qdk'\frac{\rho(k')}{(k-k')^2+c_B^2}=\frac{1}{2\pi}.
\end{equation}
Similarly $q$ is the Fermi momentum of the ground state \footnote{Interacting one dimensional system has the interesting properties that both bosons and fermions obey the exclusion principle: identical particles cannot have the same momentum, otherwise the wave function will vanish. For a detailed discussion, see Ref. [25].}, and it is determined by the density $n_B=N/L$ via the relation:
\begin{equation}
n_B=\int_{-q}^qdk~\rho(k).
\end{equation}
Now it is clear from Eqs. (\ref{eq:IntYG}) and (\ref{eq:IntLL}) that we can formally connect the Yang-Gaudin model and the Lieb-Liniger model by just a change of sign of the coupling constant $c$, where the spinless boson in the Lieb-Liniger model is identified with the singlet bound pair in the Yang-Gaudin model (thus the requirement that $m_B=2m_F$ and $n_B=2n_F$).

Now we consider the model defined in \req{eq:bosonicmodel}. It can be intuitively understood as follows: The two-component bosonic atoms denoted by $\hat{\psi}$ live on hyperplanes corresponding to different ordering sectors $X$, where $x_{X_1}<x_{X_2}<\cdots<x_{X_N}$. Without resonant couplings as in the case of the Yang-Gaudin and the Lieb-Liniger model, we only need to require the continuity of wave functions when these hyperplanes intersect. With Feshbach resonances as in our model, the molecules denoted by $\hat{\Pi}$ can be viewed as living on the intersections of the hyperplanes, which play the role of boundary conditions. As a result, we can describe the system with atomic $S$-matrices, where the molecules only enter as appropriate boundary conditions for the atomic wave functions. Next, we calculate the corresponding $S$-matrix elements to check the integrability of the present model.

In the triplet channel for $\psi$-bosons, the $\Pi$-boson is not excited, and we have both configuration $(\uparrow\uparrow)$ and $(\uparrow\downarrow+\downarrow\uparrow)/\sqrt{2}$. The two-atom states for them are
\begin{equation}
	\begin{split}
		& \ket{\text{two atom}}^{\text{tri},1}=\int dx_1dx_2~\phi(x_1,x_2)\hat{\psi}^{\dagger}_{\uparrow}(x_1)\hat{\psi}^{\dagger}_{\uparrow}(x_2)\ket{0},\\
		& \ket{\text{two atom}}^{\text{tri},0}=\\
		& \int dx_1dx_2~\phi(x_1,x_2)\frac{\hat{\psi}^{\dagger}_{\uparrow}(x_1)\hat{\psi}^{\dagger}_{\downarrow}(x_2)+\hat{\psi}^{\dagger}_{\uparrow}(x_2)\hat{\psi}^{\dagger}_{\downarrow}(x_1)}{2}\ket{0}.
	\end{split}
\end{equation}
By acting the $S$-matrix in \req{eq:Smatrix} on both wave functions, we obtain
\begin{equation}
\begin{split}
	&S(k)\ket{\text{two atom}}^{\text{tri},1}=a(k)\ket{\text{two atom}}^{\text{tri},1},\\
	&S(k)\ket{\text{two atom}}^{\text{tri},0}=\Big{(} b(k)+c(k)\Big{)}\ket{\text{two atom}}^{\text{tri},0},
\end{split}
\end{equation}
which leads to the result that $a(k)=b(k)+c(k)$. Taking into account that the spatial part of the wave functions has even parity, we can make the ansatz for $\phi(x_1,x_2)=\phi_{\text{tri}}(x_2-x_1)$ such that:
\begin{equation}
\label{eq:triplet11}
\phi_{\text{tri}}(x)=\cos(kx/2+\delta_{\text{tri}}\text{sgn}x),
\end{equation}
where $\delta_{\text{tri}}$ is the scattering phase shift. The corresponding Schr$\ddot{\text{o}}$dinger equation for $\phi_{\text{tri}}(x)$ can then be derived from the model Hamiltonian in \req{eq:bosonicmodel}:
\begin{equation}
\label{eq:triplet22}
(-2\partial^2_x-k^2/2)\phi_{\text{tri}}(x)+2g\delta(x)\phi_{\text{tri}}(0)=0,
\end{equation}
where $\phi_{\text{tri}}(0)\equiv [\phi_{\text{tri}}(0^+)+\phi_{\text{tri}}(0^-)]/2=\cos\delta_{\text{tri}}$. Integration of \req{eq:triplet22} around $x=0$ gives out
\begin{equation}
\label{eq:triplet1}
a(k)=b(k)+c(k)=e^{i2\delta_{\text{tri}}}=\frac{k-ig}{k+ig}.
\end{equation}

In the singlet channel for $\psi$-bosons, the $\Pi$-boson is also excited. The general form of the two-atom state in this case can be expressed as:
\begin{equation}
\begin{split}
	& \ket{\text{two atom}}^{\text{sin}}=\int dy ~\Phi(y)\hat{\Pi}^{\dagger}(y)\ket{0}\\
	& +\int dx_1dx_2~\phi(x_1,x_2)\frac{\hat{\psi}^{\dagger}_{\uparrow}(x_1)\hat{\psi}^{\dagger}_{\downarrow}(x_2)-\hat{\psi}^{\dagger}_{\uparrow}(x_2)\hat{\psi}^{\dagger}_{\downarrow}(x_1)}{2}\ket{0}.
\end{split}
\end{equation}
Based on symmetry considerations, the following ansatz for $\phi(x_1,x_2)$ and $\Phi(y)$ is appropriate:
\begin{equation}
\phi(x_1,x_2)=\phi_{\text{sin}}(x_2-x_1)e^{iK(x_1+x_2)}, ~~ \Phi(y)=\Phi ~e^{2iKy},
\end{equation}
where $K$ is the momentum of the mass center and the singlet wave function has odd parity
\begin{equation}
\phi_{\text{sin}}(x)=\sin(kx/2+\delta_{\text{sin}}\text{sgn}x).
\end{equation}
Then the corresponding Schr$\ddot{\text{o}}$dinger equation is:
\begin{equation}
\begin{split}
& (-2\partial^2_x-k^2/2)\phi_{\text{sin}}(x)+t^*\delta'(x)\Phi=0, \\
& -\epsilon_0\Phi-t\partial_{x}\phi(x)\Big{|}_{x\to 0^+}=k^2\Phi/2,
\end{split}
\end{equation}
which gives the expression for the $S$-matrix element in the singlet channel
\begin{equation}
\label{eq:singlet}
b(k)-c(k)=e^{2i\delta_{\text{sin}}}=\frac{4(k^2+2\epsilon_0)-i|t|^2k}{4(k^2+2\epsilon_0)+i|t|^2k}.
\end{equation}
Equation (\ref{eq:singlet}) together with \req{eq:triplet1} give us the expression of matrix elements specified in \req{eq:Smatrix}. Substituting them into the requirement in \req{eq:YangBaxter}, we finally arrive at the following condition for integrability of the present model defined in \req{eq:bosonicmodel}:
\begin{equation}
\label{eq:finetune1}
2\epsilon_0=g(g-|t|^2/4).
\end{equation}
This can be achieved by fine tuning the resonant energy $\epsilon_0$. We further introduce new coupling constants $c_1,c_2$ as:
\begin{equation}
c_1=g, ~~~ c_2=2\epsilon_0/g,
\end{equation}
such that the $S$-matrix elements take the simple form
\begin{equation}
\label{eq:ab}
a(k)=\frac{k-ic_1}{k+ic_1}, ~~~ \frac{b(k)}{a(k)}=\frac{k}{k-ic_2}.
\end{equation}
In the case with repulsive intraspecies coupling $c_1>0$, the sign of $c_2$ is controlled by the resonant energy $\epsilon_0$. The choice $c_2>0$ introduces a interspecies attraction resulting a physical pole in the singlet channel, which is exactly the singlet bound state with binding energy $\epsilon_0>0$ that we have in mind when proposing the model. Furthermore from \req{eq:finetune1} we can infer the competition between the two coupling constants:
\begin{equation}
c_1-c_2=|t|^2/4>0.
\end{equation}

Above we only considered two-body scatterings defined by the Hamiltonian in \req{eq:bosonicmodel} and we showed that the model can be made integrable by fine-tuning the resonant energy if only two-body scatterings are important. Strictly speaking, this is not sufficient for the integrability when we consider three-body and four-body scatterings (in particular, the resulting wave functions for the three-atom sector have discontinuities when a $\Pi$ particle sits right on top of a $\psi$ particle). In another word, the $\Pi$-$\psi$ and $\Pi$-$\Pi$ scatterings cannot be factorized into successive two-body scatterings. Practically, at least for relatively small densities of the system, the effect of three-body and four-body scatterings is negligible compared with two-body scatterings, since $\Pi$ particles live only on the  intersections of the hyperplanes and their scatterings are of measure zero. This is in the spirit of the so-called asymptotic Bethe ansatz \cite{Sutherland_2004,PhysRevA.73.033612,PhysRevLett.104.040402}. It is also possible to save the factorizability by introducing counterterms to the quadratic spectrum, as what has been done for the exact solution to the multichannel Kondo model \cite{PhysRevB.28.4825,PhysRevLett.52.364}. Although this recipe can be made local and Galilean invariant, it has the problem of producing a continuum of bound states when the resonance is turned off (that means $t=0$), and we don't know what will become of these bound states once the resonance is turned on. Here we disregard all these complexities and use the $S$-matrix specified in \req{eq:ab} to construct the Bethe ansatz. We believe this captures the essential features of the model as the corresponding Bethe ansatz equations give out physically sensible results. We hope to return to the issue of factorizability and resolve it in a future publication.

Before entering into the standard process of quantum inverse scattering method of solving the bosonic model, we also check the integrability of the fermionic model described in \req{eq:fermionicmodel} on the two-body scattering level, disregarding the complexities discussed above. Then the same $S$-matrix elements in \req{eq:ab} are obtained for the fermionic model but with the possibility of $c_1<c_2$. We follow the same procedure as the bosonic case. The corresponding Schr$\ddot{\text{o}}$dinger equation in the triplet channel is then
\begin{equation}
\begin{split}
& (-2\partial^2_{x}-k^2/2)\phi_{\text{tri}}(x)+t^*_{\Xi}\delta'(x)\Phi_{\Xi}=0, \\
& -\epsilon_{\Xi} \Phi_{\Xi}-t_{\Xi}\partial_{x}\phi_{\text{tri}}(x)_{x\to 0^+}=k^2\Phi_{\Xi}/2,
\end{split}
\end{equation}
where $\phi_{\text{tri}}(x)=\sin(kx/2+\delta_{\text{tri}}\text{sgn}x)$ is the spatial part of the triplet wave function and $\Phi_{\Xi}$ is the amplitude of the vector resonance wave function. We then obtain the $S$-matrix element in the triplet channel:
\begin{equation}
\small
\label{eq:tri}
a(k)=b(k)+c(k)=e^{2i\delta_{\text{tri}}}=\frac{4(k^2+2\epsilon_{\Xi})-i|t_{\Xi}|^2k}{4(k^2+2\epsilon_{\Xi})+i|t_{\Xi}|^2k}.
\end{equation}
Similarly, in the singlet channel we have the following Schr$\ddot{\text{o}}$dinger equation:
\begin{equation}
\begin{split}
& (-2\partial^2_x-k^2/2)\phi_{\text{sin}}(x)+[2g\phi_{\text{sin}}(0)+t^*_{\Pi}\Phi_{\Pi}]\delta(x)=0, \\
& -\epsilon_{\Pi} \Phi_{\Pi}+t_{\Pi}\phi_{\text{sin}}(0)=k^2\Phi_{\Pi}/2,
\end{split}
\end{equation}
where $\phi_{\text{sin}}(x)=\cos(kx/2+\delta_{\text{sin}}\text{sgn}x)$ is the spatial part of the singlet wave function and $\Phi_{\Pi}$ is the amplitude of the scalar resonance wave function. We then obtain the $S$-matrix element in the singlet channel:
\begin{equation}
\small
\label{eq:sin}
b(k)-c(k)=e^{2i\delta_{\text{sin}}}=\frac{k(k^2+2\epsilon_{\Pi})-i[g k^2+2g\epsilon_{\Pi}+|t_{\Pi}|^2]}{k(k^2+2\epsilon_{\Pi})+i[g k^2+2g\epsilon_{\Pi}+|t_{\Pi}|^2]}.
\end{equation}

Equation (\ref{eq:tri}) together with \req{eq:sin} produce the same $S$-matrix elements in \req{eq:ab} if the following fine tuning is applied:
\begin{equation}
\epsilon_{\Xi}=0, ~~~ \frac{|t_{\Xi}|^2}{4}\left(\frac{|t_{\Xi}|^2}{4}-g\right)=-\frac{|t_{\Pi}|^2}{g}=2\epsilon_{\Pi},
\end{equation}
and the coupling constants $c_1,c_2$ are defined as:
\begin{equation}
\label{eq:collapse}
c_1=|t_{\Xi}|^2/4, ~~~ c_1-c_2=-|t_{\Pi}|^2/(2\epsilon_{\Pi}).
\end{equation}
If the scalar resonance is made a singlet bound state with binding energy $\epsilon_{\Pi}>0$, then we realize the possibility of $c_1<c_2$ mentioned above.

Since both the bosonic and fermionic model present the same form of $S$-matrix elements in \req{eq:ab}, we can discuss their exact solutions together by substituting \req{eq:ab} into \req{eq:Nested}. The resulting algebraic Bethe ansatz equations in the sector where $M$ out of $N$ atoms are placed spin-down are:
\begin{equation}
\label{eq:Bethe}
\begin{split}
& (-)\prod_{n=1}^N\frac{\Lambda_{\alpha}-k_n-ic'_2}{\Lambda_{\alpha}-k_n+ic'_2}=\prod_{\beta=1}^M\frac{\Lambda_{\alpha}-\Lambda_{\beta}-ic_2}{\Lambda_{\alpha}-\Lambda_{\beta}+ic_2},\\
& (\mp)^{N-1}e^{-ik_jL}=\prod_{n=1}^N\frac{k_j-k_n-ic_1}{k_j-k_n+ic_1}\prod_{\alpha=1}^M\frac{\Lambda_{\alpha}-k_j-ic'_2}{\Lambda_{\alpha}-k_j+ic'_2},
\end{split}
\end{equation}
where $L$ is the size of the system, $c'_2=c_2/2$, $j,n=1,2,\cdots,N$ and $\alpha,\beta=1,2,\cdots,M$. The conventional shift $\Lambda_{\alpha}\to \Lambda_{\alpha}+ic'_2$ is introduced. Also the upper sign is for bosons, and the lower sign is for fermions. In the following sections we discuss the two cases with $c_1>c_2>0$ and $c_2>c_1>0$ respectively.

\section{Uniform Regime with $c_1>c_2>0$}
\label{sec:uniform}
In this section, we discuss the present model in the parameter regime $c_1>c_2>0$, where the ground state is a Fermi sea of the two-atom bound pair. Firstly, we derive the equation for the density of states in the thermodynamic limit, where we analyze its singular behavior due to level condensation and find that the Fermi momentum is bounded from above. Then we discuss the physics of BCS-BEC crossover in the context of the present model, where we find that the extremes of the excitation spectra have robust features and the system develops a collapsing instability in the limit $c_2\to c_1$. After that, we discuss the zero temperature phase diagram of the present model with an external magnetic field, of which a considerable part is occupied by the one dimensional analog of the FFLO state. Besides, a critical magnetic field arises due to the presence of the upper bound for the Fermi momentum. This is in contrast to the situation for the Yang-Gaudin model, where an arbitrarily small external magnetic field can polarize the ground state as long as the particle density of the system is large enough.

\subsection{Level Condensation and Limiting Fermi Momentum $Q^*$}
Let us start with the system with even number of atoms. Then the ground state is a Fermi sea of two-strings with the same structure as in the Yang-Gaudin model:
\begin{equation}
\label{eq:twostring}
k_{\alpha,1}=\Lambda_{\alpha}+ic'_2, ~~~ k_{\alpha,2}=\Lambda_{\alpha}-ic'_2,
\end{equation}
where $\alpha=1,2,\cdots,M=N/2$ and the Bethe ansatz equations in \req{eq:Bethe} reduce to the equations for the center momenta $\Lambda_{\alpha}$:
\begin{equation}
\label{eq:Bethe2}
\begin{split}
& e^{i2\Lambda_{\alpha}L}= (-)\prod_{\beta=1}^{M}\left(\frac{\Lambda_{\alpha}-\Lambda_{\beta}-ic_2}{\Lambda_{\alpha}-\Lambda_{\beta}+ic_2}\right)\left(\frac{\Lambda_{\alpha}-\Lambda_{\beta}+ic_1}{\Lambda_{\alpha}-\Lambda_{\beta}-ic_1}\right)^2\\
& \left(\frac{\Lambda_{\alpha}-\Lambda_{\beta}+i(c_1+c_2)}{\Lambda_{\alpha}-\Lambda_{\beta}-i(c_1+c_2)}\right)\left(\frac{\Lambda_{\alpha}-\Lambda_{\beta}+i(c_1-c_2)}{\Lambda_{\alpha}-\Lambda_{\beta}-i(c_1-c_2)}\right).
\end{split}
\end{equation}
Because the center momenta $\Lambda_{\alpha}$ in the ground state are all real, we can take the logarithm of the above equations:
\begin{equation}
\label{eq:discrete}
2\Lambda_{\alpha}L=2\pi J_{\alpha}-\sum_{\beta=1}^M\theta(\Lambda_{\alpha}-\Lambda_{\beta}),
\end{equation}
where $J_{\alpha}$s are consecutive integers or half-odd integers depending on:
\begin{equation}
J_{\alpha}=\frac{M+1-2\alpha}{2}, ~~~ \alpha=1,2,\cdots,M.
\end{equation}
The phase-shfit function $\theta(\Lambda-\Lambda')$ is defined as:
\begin{equation}
\label{eq:smooth}
\begin{split}
\theta(\Lambda-&\Lambda')=-2\arctan\left(\frac{\Lambda-\Lambda'}{c_2} \right)+4\arctan\left( \frac{\Lambda-\Lambda'}{c_1} \right)\\
& +2\arctan\left( \frac{\Lambda-\Lambda'}{c_1+c_2} \right) +2\arctan\left( \frac{\Lambda-\Lambda'}{c_1-c_2} \right).
\end{split}
\end{equation}
We can see that \req{eq:smooth} reduces to \req{eq:PhaseYG} in the limit $c_2\ll 1\ll c_1$, and to \req{eq:PhaseLL} in the limit $0<c_1-c_2\ll 1 \ll c_1$, so the present model has the Yang-Gaudin model and the Lieb-Liniger model as its two limiting models. Also the total energy and momentum of the system can then be expressed as:
\begin{equation}
\label{eq:discrete2}
E=\sum_{\alpha=1}^M\left[\left(\Lambda_{\alpha}+\frac{ic_2}{2} \right)^2+\left(\Lambda_{\alpha}-\frac{ic_2}{2} \right)^2 \right], ~ P=\sum_{\alpha=1}^M2\Lambda_{\alpha}.
\end{equation}

In the thermodynamic limit, \req{eq:discrete} can be replaced by an integral equation for the density of state $\sigma(\Lambda_j)=1/[L(\Lambda_{j+1}-\Lambda_j)]$, just like what we discussed previously:
\begin{equation}
\label{eq:kernel0}
\begin{split}
& \sigma(\Lambda)-\int_{-Q}^Qd\Lambda'~K(\Lambda-\Lambda')\sigma(\Lambda')=\frac{1}{\pi},\\
& K(\Lambda-\Lambda')=\frac{1}{2\pi}\theta'(\Lambda-\Lambda'),
\end{split}
\end{equation}
where $Q$ is the Fermi momentum of the ground state, and it is determined by the density $n=N/L$ via the relation:
\begin{equation}
\label{eq:density}
\frac{n}{2}=\int_{-Q}^Qd\Lambda~\sigma(\Lambda).
\end{equation}
Also the total energy and momentum of the system in the thermodynamic limit can be expressed using the density of state $\sigma(\Lambda)$:
\begin{equation}
	E=\int_{-Q}^Qd\Lambda\left(2\Lambda^2-\frac{c^2_2}{2} \right)\sigma(\Lambda), P=\int_{-Q}^Qd\Lambda~2\Lambda\sigma(\Lambda).
\end{equation}

For further analysis of \req{eq:kernel0}, it is useful to rescale the parameters with respect to $Q$:
\begin{equation}
\label{eq:scaled}
x\equiv \frac{\Lambda}{Q}, ~~~ \lambda\equiv \frac{c_1}{Q}, ~~~ \xi\equiv \frac{c_2}{c_1}.
\end{equation}
Correspondingly, the integral equation for the density of states with the rescaled parameters is:
\begin{equation}
\label{eq:thermo}
	\sigma(x)-\int_{-1}^1dx'~K(x-x')\sigma(x')=\frac{1}{\pi},
\end{equation}
where $K(x)$ expressed in the rescaled parameters is
\begin{equation}
\label{eq:kk}
	\begin{split}
		& K(x)=\frac{1}{\pi}\left(-\frac{\xi\lambda}{x^2+(\xi\lambda)^2}+\frac{2\lambda}{x^2+\lambda^2}\right.\\
	&~~~~~~\left.+\frac{(1+\xi)\lambda}{x^2+(1+\xi)^2\lambda^2}+\frac{(1-\xi)\lambda}{x^2+(1-\xi)^2\lambda^2)} \right),
	\end{split}
\end{equation}
and \req{eq:density} is equivalent to
\begin{equation}
\label{eq:density2}
	\int_{-1}^1dx~\sigma(x)=\frac{n}{2Q}.
\end{equation}
By construction, the density of states $\sigma(x)\geqslant 0$ for $x\in [-1,1]$. Correspondingly, the kernel
\begin{equation}
\label{eq:kernel}
	\mathcal{L}(x,x')=\delta(x-x')-K(x-x')
\end{equation}
in \req{eq:thermo} must also be a positive definite operator, with its smallest eigenvalue bigger than $1/(\pi\sigma_{\text{max}})$, where $\sigma_{\text{max}}$ is the maximum value of $\sigma(x)$ on the interval $x\in[-1,1]$ \cite{Korepin_1993}. This is indeed the case in the Lieb-Liniger model and the Yang-Gaudin model for any coupling strength.

\begin{figure}[htp!]
	\includegraphics[scale=0.15]{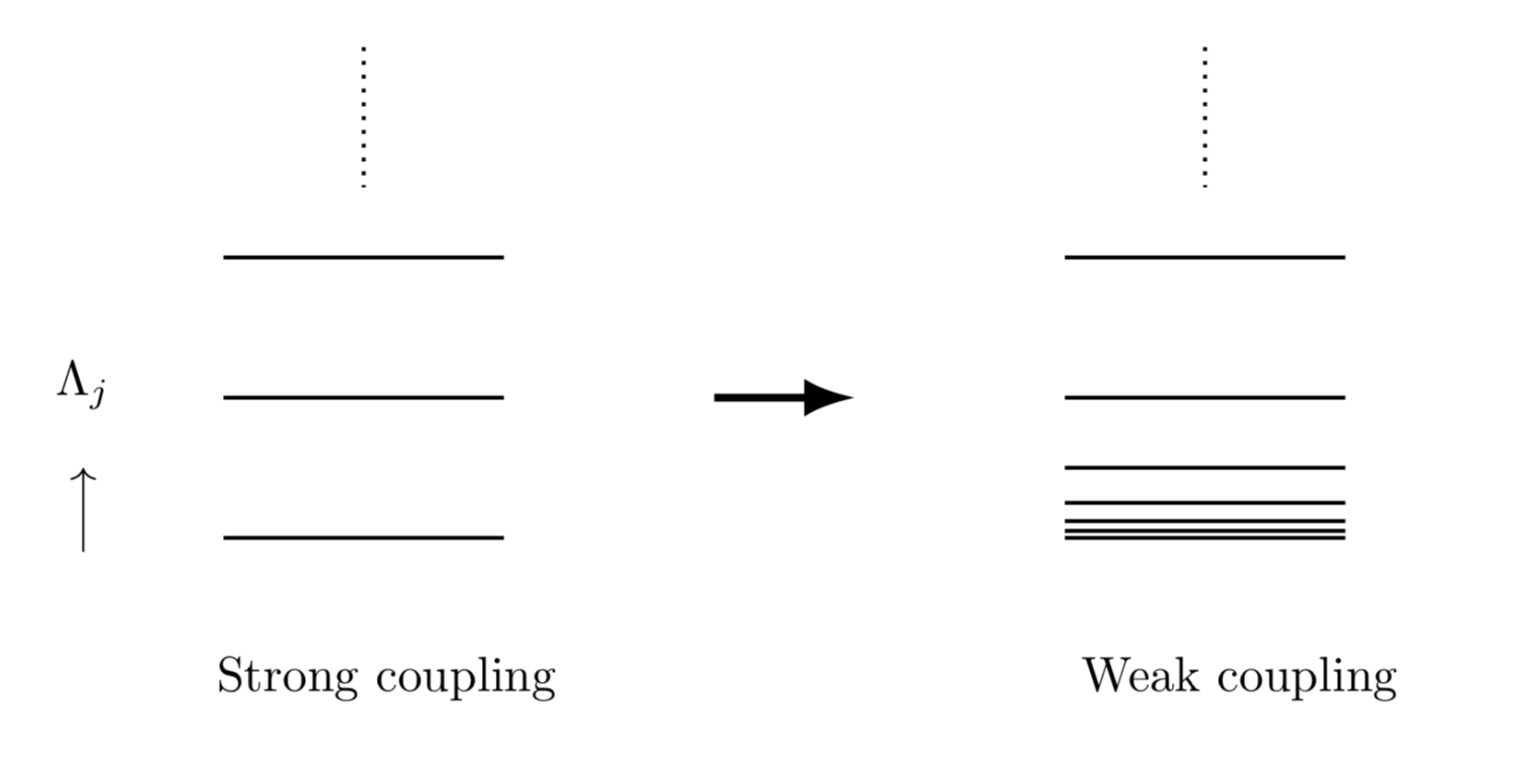}
	\caption{\footnotesize Condensation of levels for weakly interacting bosons in one dimension. Left pane is drawn for strong coupling and right pane is draw for weak coupling.}
	\label{fig:coupling}
\end{figure}

However, in the present model, the kernel $\mathcal{L}(x,x')$ in \req{eq:kernel} is positive definite only for $\lambda>\lambda^*=\lambda^*(\xi)$, where $\lambda^*(\xi)$ will be found later. Equivalently, \req{eq:scaled} yields an upper bound for the Fermi momentum:
\begin{equation}
	Q<Q^*=c_1/\lambda^*.
\end{equation}
Then, the apparent conflict appears: How do we satisfy the conservation of the mass density of the system in Eqs. (\ref{eq:density}) and (\ref{eq:density2}) while the Fermi momentum $Q$ is limited by $Q^*$? Solution of this problem is to have $\sigma(x)\to \infty$ at $n\to \infty$ while keeping $Q^*$ fixed. The only way to have such a result is to require the operator $\mathcal{L}(x,x')$ to have an almost zero mode at $\lambda\to \lambda^*+0$:
\begin{equation}
\label{eq:zero}
	\int_{-1}^1dx'\mathcal{L}_{\lambda^*}(x,x')\sigma^0(x)=0.
\end{equation}
This phenomenon reminds us the condensation of levels which is characteristic for weakly interacting bosons (see Fig. \ref{fig:coupling}).

\begin{figure}
\includegraphics[scale=0.4]{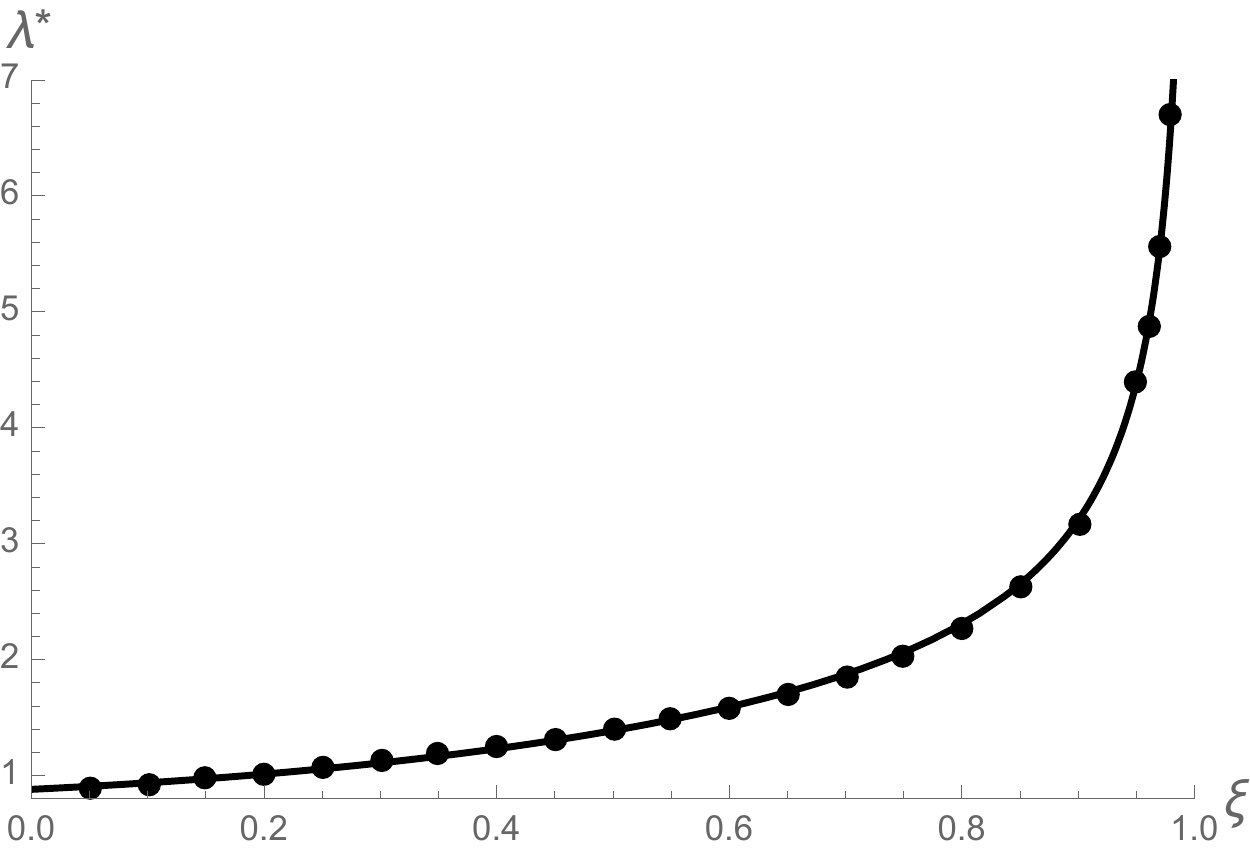}
\caption{\footnotesize The critical value $\lambda^*$ as a function of $\xi=c_2/c_1$. The dots are obtained numerically and the thick line is the curve of the fitting function in \req{eq:fit}. We can see that as $\xi\to1$, $\lambda^*\to \infty$, which means that $Q^*$ tends to zero, this implies the instability of the system.}
\label{fig:critical}
\end{figure}

For a fixed value of $\xi$ and different values of $\lambda$, we solve \req{eq:thermo} numerically using the modified quadrature method \cite{Rahbar2008ACA}, where the usual quadrature approximation to the integral is modified to give more accurate result for a weakly singular kernel. The critical value $\lambda^*(\xi)$ is determined by the point at which the value of the solution $\sigma(x)$ to \req{eq:thermo} changes from positive to negative. Accordingly, for a fixed value of $\xi$, the zero mode $\sigma^0(x)$ for $\mathcal{L}_{\lambda^*}(x,x')$ can be approximated by the singular part of the solution $\sigma(x)$ that grows with increasing density $n$ when $\lambda\to \lambda^*(\xi)+0$. The numerically determined critical curve $\lambda^*(\xi)$ is shown in Fig. \ref{fig:critical},
and it can be fitted by the formula:
\begin{equation}
\label{eq:fit}
	\lambda^*(\xi)=\sqrt{\frac{0.75}{1-\xi}}+0.015+0.045\xi+0.50\xi^2, ~~~ \xi\in (0,1),
\end{equation}
which has the limiting behavior that $\lambda^*(\xi\to 0)\approx 0.88$ and $\lambda^*(\xi\to 1)\to \infty$. In the latter limit, the upper bound $Q^*$ goes to zero and the system develops a collapsing instability which we will discuss in Sec. \ref{sec:string}.

\begin{figure}[htp!]
	\includegraphics[scale=0.13]{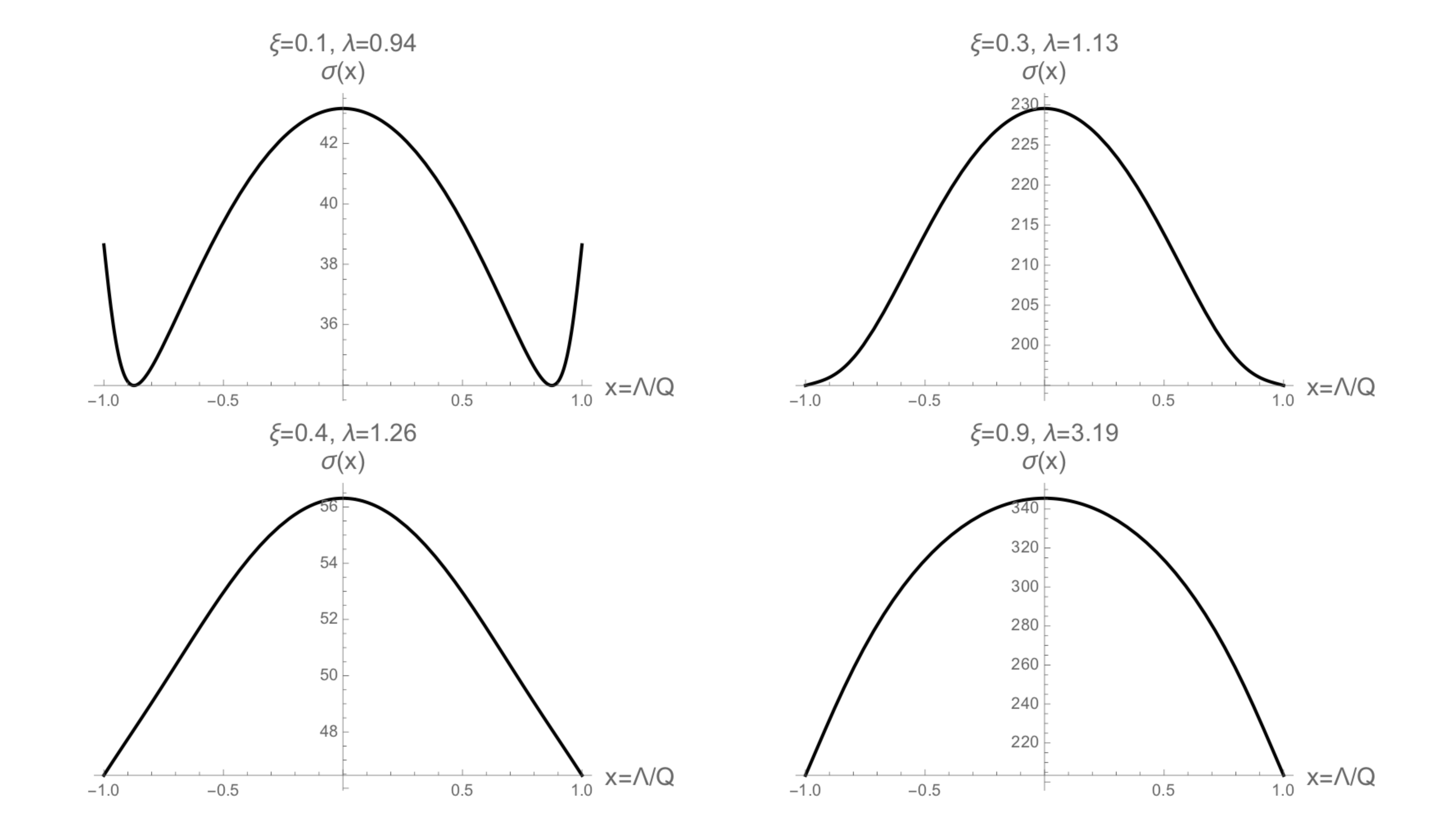}
	\caption{\footnotesize The solution $\sigma(x)$ to \req{eq:thermo} for $\lambda\to \lambda^*(\xi)+0$. $\sigma(x)$ can be separated into the regular part $\sigma_{\text{reg}}(x)$ and the singular part $\sigma^0(x)$, where the latter grows with increasing $n$ and is seen in the figure as the part near $x=0$.}
	\label{fig:critical2}
\end{figure}

The solution $\sigma(x)$ to \req{eq:thermo} at $\lambda\to\lambda^*+0$ for several values of $\xi$ is shown in Fig. \ref{fig:critical2}, and it consists of a regular and a singular part:
\begin{equation}
	\sigma(x)=\sigma_{\text{reg}}(x)+\sigma^0(x),
\end{equation}
where the singular part $\sigma^0(x)$ can be viewed as the approximated zero mode defined in \req{eq:zero}, thus we use the same symbol for them.

In the limit $\xi\to 1$, we can extract the analytical expression for the zero mode from \req{eq:thermo}. For $\xi\to 1$, we have $(1-\xi)\lambda^*(\xi)\to 0$ from \req{eq:fit}, then $K(x-x')$ in \req{eq:kk} at $\lambda\to \lambda^*+0$ can be approximated as
\begin{equation}
\label{eq:approx4}
	\begin{split}
		K(x-x')&\approx \frac{1}{\pi}\left(\frac{3}{2\lambda}+\frac{(1-\xi)\lambda}{(x-x')^2+(1-\xi)^2{\lambda}^2} \right), \\
		&\approx \frac{3}{2\pi\lambda}+\delta(x-x')-\frac{(1-\xi)\lambda}{\pi}\frac{\partial}{\partial x}\mathcal{P}\frac{1}{x-x'},
	\end{split}
\end{equation}
where the symbol $\mathcal{P}$ represents the Cauchy principle part. As a result, \req{eq:thermo} can be reduced to an equation for the zero mode $\sigma^0(x)$:
\begin{equation}
\label{eq:newreduced}
	-\frac{3}{2\pi\lambda}\int_{-1}^1dx'\sigma^0(x')+\frac{(1-\xi)\lambda}{\pi}\frac{\partial}{\partial x}\mathcal{P}\int_{-1}^1dx'\frac{\sigma^0(x')}{x-x'}=\frac{1}{\pi}.
\end{equation}
Using the identity that
\begin{equation}
	\mathcal{P}\int_{-1}^1\frac{dx'}{x-x'} \sqrt{1-{x'}^2}=\pi x,
\end{equation}
the solution $\sigma^0(x)$ to \req{eq:newreduced} can be determined as
\begin{equation}
\label{eq:zeromode}
	\sigma^0(x)=\frac{4\lambda}{\pi}\frac{1}{4(1-\xi){\lambda}^2-3}\sqrt{1-x^2},
\end{equation}
where the value of $\lambda$ is then fixed by the normalization condition in \req{eq:density2}:
\begin{equation}
\label{eq:norm4}
	\int_{-1}^1dx'\sigma^0(x')=\frac{n}{2Q}=\frac{n}{2c_1}\lambda.
\end{equation}
The critical value $\lambda^*$ is given by either the condition that $\sigma^0(x)\geqslant 0$ or by \req{eq:norm4} in the limit $n\to \infty$:
\begin{equation}
\label{eq:qtoqstar}
	\lambda^*=\sqrt{\frac{3}{4(1-\xi)}},
\end{equation}
which agrees with the fitting formula in \req{eq:fit} in the limit $\xi\to 1$. Using \req{eq:qtoqstar}, we can rewrite $\sigma^0(x)$ in \req{eq:zeromode} as
\begin{equation}
\label{eq:zeromode2}
	\sigma^0(x)=\frac{4\lambda{\lambda^*}^2}{3\pi}\frac{\sqrt{1-x^2}}{\lambda^2-{\lambda^*}^2}= \frac{2c_1}{3\pi}\frac{\sqrt{1-x^2}}{Q^*-Q},
\end{equation}
where $Q=c_1/\lambda$ and $Q^*=c_1/\lambda^*$. This square root singularity in $\sigma^0(x)$ also appears in the Lieb-Liniger model when the coupling constant $c_B$ approaches zero from the positive side \cite{PhysRev.130.1605,hutson_1963}. Here it appears in the limit $\xi\to 1$ when $Q$ approaches $Q^*$ from below. For $\xi$ smaller than $1$, the zero mode $\sigma^0(x)$ acquires correction to the form in \req{eq:zeromode2} near the boundaries $x=\pm 1$ (see Fig. \ref{fig:critical2}). Numerical calculation shows that this correction is negligible up to the point $\xi=0.4$, so the functional form in \req{eq:zeromode2} provides a good description of the zero mode $\sigma^0(x)$ in the range $0.4<\xi<1$, and we only need to replace the prefactor $2c_1/(3\pi)$ with a positive function $F(Q^*)$ that depends on $Q^*$.

Far away from the lower bound $\lambda^*$, asymptotic behaviors for the density of state $\sigma(x)$ can be extracted from \req{eq:thermo} in the limit $\lambda\to \infty$ with $\lim_{\xi\to 0}\xi\lambda\to 0$ and $\lim_{\xi\to 1}(1-\xi)\lambda\to 0$. For $\xi\to 0$, \req{eq:thermo} reduces to
\begin{equation}
\label{eq:YGddd}
	\sigma(x)+\frac{1}{\pi}\int_{-1}^1dx'\frac{\xi\lambda}{(x-x')^2+(\xi\lambda)^2}\sigma(x')=\frac{1}{\pi},
\end{equation}
which coincides \req{eq:IntYG} for the Yang-Gaudin model if we rescale the parameters there accordingly. The asymptotic behavior for the solution $\sigma(x)$ to \req{eq:YGddd} in the limit $\xi\lambda\to 0$ has already been obtained in literature \cite{gaudin_2014,doi:10.1143/JPSJ.74.1724}:
\begin{equation}
\label{eq:YGsigma}
	\sigma(x)=\frac{1}{2\pi}\left(1+\frac{\xi\lambda}{\pi}\frac{1}{1-x^2}+\cdots \right).
\end{equation}
For $\xi\to 1$, \req{eq:thermo} reduces to
\begin{equation}
\label{eq:LLddd}
	\sigma(x)-\frac{1}{\pi}\int_{-1}^1dx'\frac{(1-\xi)\lambda}{(x-x')^2+(1-\xi)^2\lambda^2}\sigma(x')=\frac{1}{\pi},
\end{equation}
which coincides \req{eq:IntLL} for the Lieb-Liniger model if we rescale the parameters there accordingly. The asymptotic behavior for the solution $\sigma(x)$ to \req{eq:LLddd} in the limit $(1-\xi)\lambda\to 0$ has also been obtained in the literature \cite{hutson_1963,gaudin_2014}:
\begin{equation}
\label{eq:LLsigma}
\begin{split}
	\sigma(x)=&\frac{1}{\pi(1-\xi)\lambda}\sqrt{1-x^2}\\
	+&\frac{1}{2\pi^2}\frac{1}{\sqrt{1-x^2}}\left[x\ln\left(\frac{1-x}{1+x}\right)+\ln\frac{16\pi e}{(1-\xi)\lambda} \right]+\cdots.
\end{split}
\end{equation}
The two asymptotic behaviors in Eqs. (\ref{eq:YGsigma}) and (\ref{eq:LLsigma}) will be of use in the next section when we calculate the asymptotic behaviors of physical observables in the BCS and BEC limits of the present model.

\subsection{BCS-BEC Crossover without External Magnetic Field}

We will now show that the present model can provide an ideal scenario for BCS-BEC crossover in one dimension, without the drawbacks of simply connecting Yang-Gaudin model with Lieb-Liniger model. The behavior of the system is controlled by two dimensionless coupling constants:
\begin{equation}
	\gamma_1=c_1/n, ~~~ \gamma_2=c_2/n,
\end{equation}
where $m_{\psi}n=m_{\psi}N/L$ is the total mass density of the system with $m_{\psi}$ being the mass of the atom. The ratio between the two dimensionless coupling constants is $\gamma_2/\gamma_1=\xi$.

We consider the situation that $\gamma_1\gg 1$ is kept fixed at a large value and $\gamma_2$ is varied such that $\xi$ goes from 0 to 1. In the limit $\xi\to 0$, the dominant term in \req{eq:smooth} is:
\begin{equation}
\theta(\Lambda-\Lambda')\approx -2\arctan\left( \frac{\Lambda-\Lambda'}{c_2} \right),
\end{equation}
which coincides with the Yang-Gaudin model (see \req{eq:PhaseYG}). Thus the BCS limit can be realized by tuning $\gamma_2$ to the limit $\xi\to 0$, where the system is weakly coupled with small dimensionless coupling $\gamma_2$. As we tune the coupling $\gamma_2$ larger, the system evolves toward the fermionic super Tonks-Girardeau gas, where the Yang-Gaudin model terminates. If we tune the coupling $\gamma_2$ even larger, the system can overcome this strong coupling limit and develop bosonic behaviors. In the limit $\xi\to 1$, the dominant term in \req{eq:smooth} is:
\begin{equation}
\label{eq:BECphase1}
\theta(\Lambda-\Lambda')\approx 2\arctan\left( \frac{\Lambda-\Lambda'}{c_1-c_2} \right),
\end{equation}
which coincides with the Lieb-Liniger model (see \req{eq:PhaseLL}). Thus the BEC limit is realized by tuning $\gamma_2$ to the limit $\xi\to 1$, where the system is again weakly coupled with small dimensionless coupling $\delta \gamma=\gamma_1-\gamma_2$. We can see that the present model describes the BCS-BEC crossover in a unified fashion, with a single Hamiltonian governing the evolution.

\begin{figure}
\includegraphics[scale=0.13]{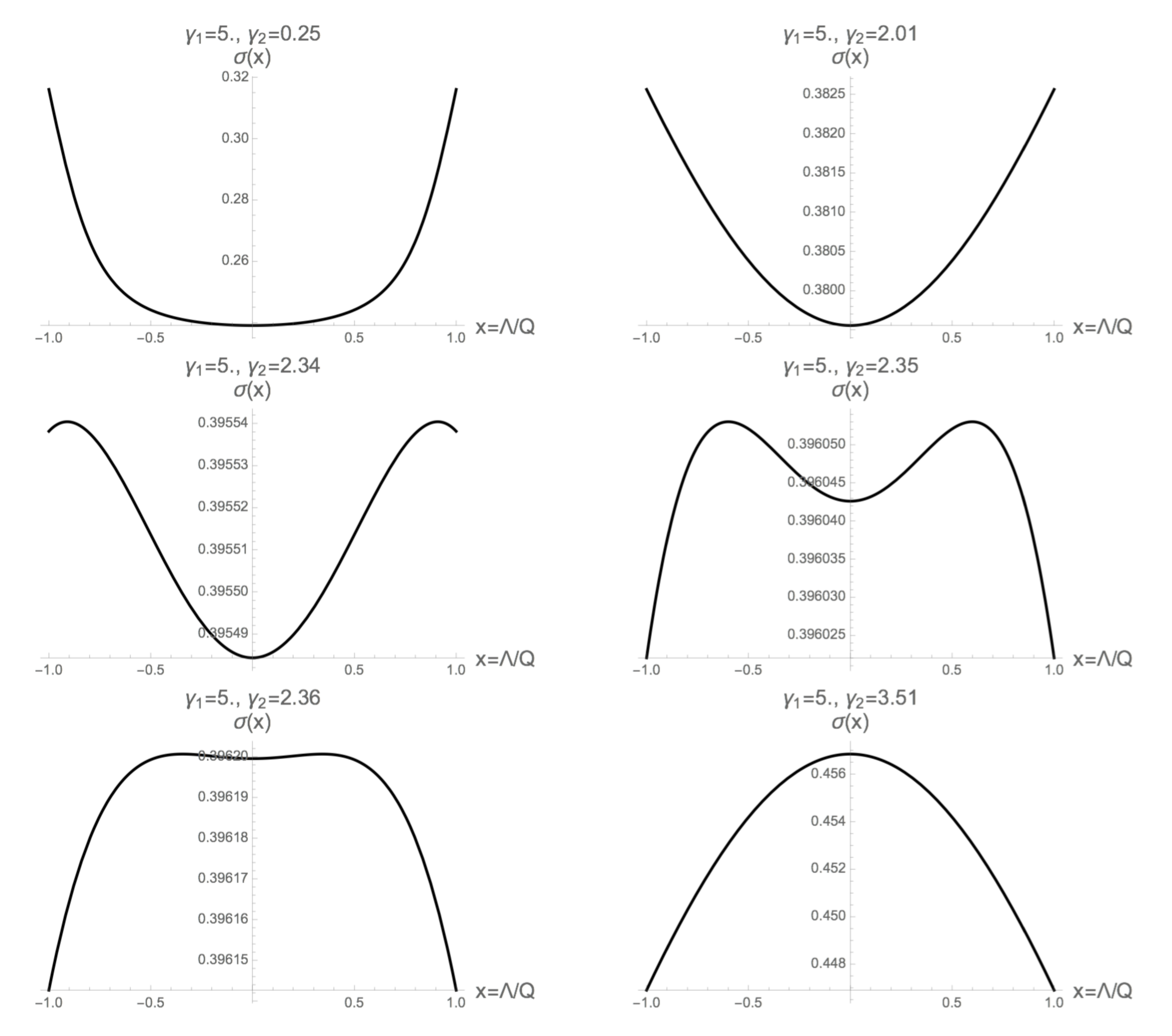}
\caption{\footnotesize Density of states $\sigma(x)$ with fixed $\gamma_1$ and varying $\gamma_2$, where $\gamma_{1,2}\equiv c_{1,2}/n$. By tuning $\gamma_2$ toward $\gamma_1$, the behavior of the system changes from weakly attractive fermions (nearly fat at the center and sharp increase near the boundary) to fermionic super Tonks-Girardeau gas and finally to weakly interacting bosons (condensation at the center).}
\label{fig:distribution}
\end{figure}

We first demonstrate the crossover between fermionic and bosonic behaviors by the evolution of density of states $\sigma(x)$ with varying $\gamma_2$. The asymptotic behaviors of $\sigma(x)$ in the BCS and BEC limits are shown in Eqs. (\ref{eq:YGsigma}) and (\ref{eq:LLsigma}), and a typical result of $\sigma(x)$ obtained by solving \req{eq:thermo} together with \req{eq:density2} numerically is shown in \reffig{fig:distribution}, where the crossover between flat distribution in BCS limit and level condensation in BEC limit is transparent.

Next, we discuss the ground state properties and low energy excitations of the present model at BCS-BEC crossover respectively.

\subsubsection{Ground State Properties and Instability}
We analyze the ground state energy and further the compressibility of the system. The ground state energy density $\epsilon$ and conserved density $n$ of the system can be calculated using the density of states $\sigma(x)$:
\begin{equation}
\label{eq:ENperL}
\begin{split}
	\epsilon=\frac{E}{L}=&\int_{-Q}^Qd\Lambda\left(2\Lambda^2-\frac{c^2_2}{2} \right)\sigma(\Lambda)\\
	=&Q^3\int_{-1}^1dx\left(2x^2-\frac{\xi^2\lambda^2}{2}\right)\sigma(x), \\
	n=\frac{N}{L}=& \int_{-Q}^Qd\Lambda~2\sigma(\Lambda)=Q\int_{-1}^1dx~2\sigma(x),
\end{split}
\end{equation}
and the compressibility $\kappa$ can be calculated from the energy density $\epsilon$ using the standard thermodynamic relation:
\begin{equation}
\label{eq:compress}
	 \frac{1}{\kappa}=n^2\left(\frac{d^2\epsilon}{dn^2}\right).
\end{equation}

The asymptotic behaviors of $\epsilon$ and $\kappa$ in the BCS ($\xi\to 0$) and BEC ($\xi\to 1$) limits are obtained by substituting Eqs. (\ref{eq:YGsigma}) and (\ref{eq:LLsigma}) into Eqs. (\ref{eq:ENperL}) and (\ref{eq:compress}):
\begin{equation}
\label{eq:asymptotic}
\begin{split}
\frac{\epsilon+nc^2_2/4}{n\epsilon_F}\stackrel{\xi\to 0}{\approx} & ~\frac{1}{3}\left(1-\frac{6\gamma_1}{\pi^2}\xi+\cdots\right) \\
\stackrel{\xi\to 1}{\approx} & ~\frac{\gamma_1(1-\xi)}{\pi^2}\left(1-\frac{8\sqrt{\gamma_1(1-\xi)}}{3\pi}+\cdots \right),\\
\frac{1}{\kappa n^3} \stackrel{\xi\to 0}{\approx} & ~\frac{\pi^2}{2}\left(1-\frac{2\gamma_1}{\pi^2}\xi+\cdots \right)\\
\stackrel{\xi\to 1}{\approx} & ~\frac{\gamma_1(1-\xi)}{2}\left(1-\frac{\sqrt{\gamma_1(1-\xi)}}{\pi}+\cdots \right).
\end{split}
\end{equation}
where $\epsilon_F=\pi^2n^2/4$ is the Fermi energy for the noninteracting Fermi gas and $c^2_2/4$ is the binding energy per atom. The results in \req{eq:asymptotic} agree with those for the Yang-Gaudin model and the Lieb-Liniger model in the weak coupling limit respectively \cite{PhysRev.130.1605,1975JETP...40..781K,PhysRevLett.93.090408}.

\begin{figure}
\includegraphics[scale=0.13]{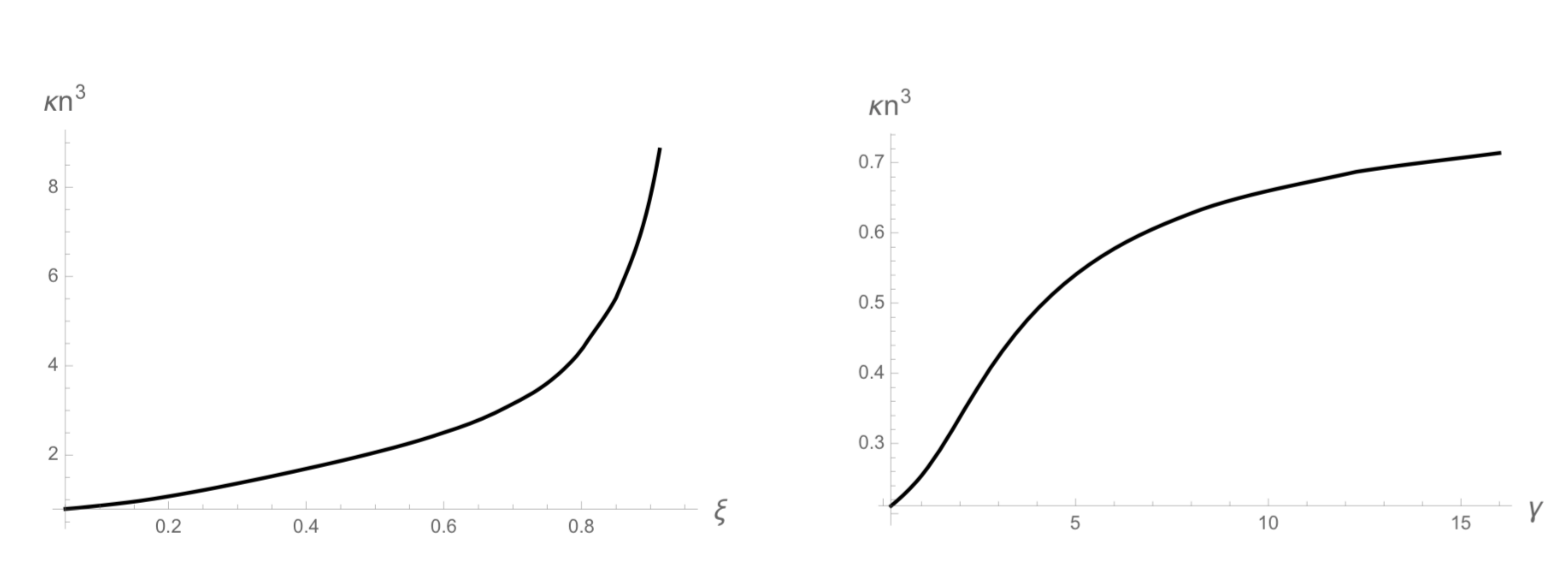}
\caption{\footnotesize Numerical result for compressibility, where we have made it dimensionless by multiplying by $n^3$. Left panel is plotted for the present model with $\gamma_1=5$, where $\kappa n^3$ changes with $\xi=c_2/c_1$. Right panel is plotted for the Yang-Gaudin model by varying dimensionless coupling constant $\gamma=c_F/n$. The difference is apparent: for the present model, there is divergence for compressibility, while for the Yang-Gaudin model, the compressibility saturates at a finite value.}
\label{fig:compress}
\end{figure}

For general values of $\xi$, we numerically solve \req{eq:thermo} for the density of states $\sigma(x)$ and numerically calculate the ground state energy and compressibility. A typical result for the compressibility is shown in \reffig{fig:compress}, together with the result for the Yang-Gaudin model for a comparison. From \reffig{fig:compress}, we can see that instead of saturating at a finite value as in the Yang-Gaudin model, the compressibility of the system in the present model diverges in the limit $\xi\to 1$, just as shown by the asymptotic behavior in \req{eq:asymptotic}. This means that the system becomes infinitely compressible and the spatially uniform ground state becomes unstable. In the meantime, the upper bound $Q^*$ for the Fermi momentum goes to zero in the same limit (see \reffig{fig:critical}). These facts signal a collapsing instability in the system, which we will discuss in the Sec. \ref{sec:string}.

\subsubsection{Excitations and Robustness of Their Extremes}
We analyze the low energy excitations of the present model. There are two types of them, classified according to the spin quantum number - the $S=0$ excitations and the $S=1/2$ excitations. We first summarize their features and then present the detailed analysis.

\begin{figure}[htp!]
\includegraphics[scale=0.5]{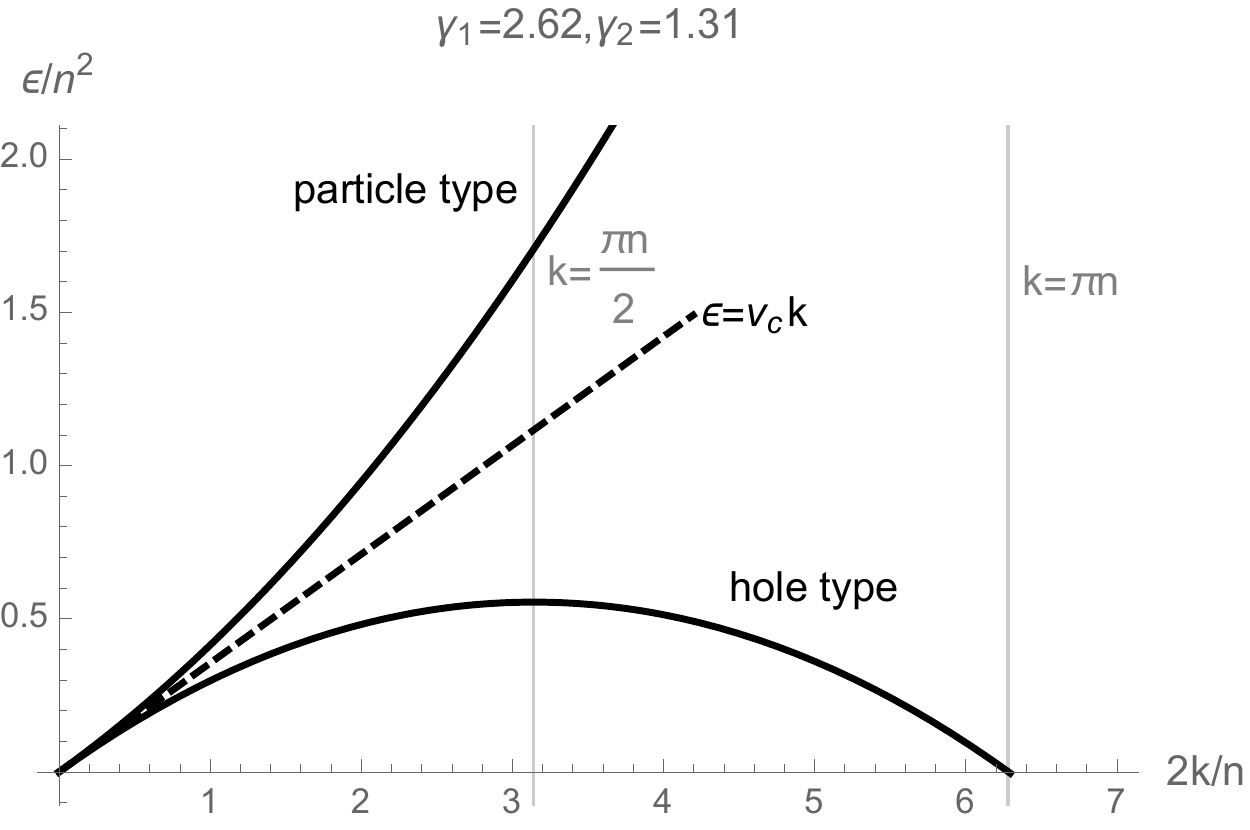}
\caption{\footnotesize A typical spectrum for the $S=0$ excitations. There are two branches, one for hole type and one for particle type. At long wavelength, they both reduce to the phonon branch. The maximum of the hole branch is fixed at $k_{\text{max}}=k_F=\pi n/2$ and there is a periodicity of $2k_F=\pi n$ in the hole branch.}
\label{fig:dispersion}
\end{figure}

The spectrum of the $S=0$ excitations in the present model has two branches, one of them is the usual Bogoliubov quasiparticle branch and the other is similar to the type-II branch in the Lieb-Liniger model \cite{PhysRev.130.1616}. Physically, the former can be identified as the particle branch and the latter can be identified as the hole branch (see \reffig{fig:dispersion}), a classification due to the fact that the structure of the ground state is the same as that of a Fermi system. There are two robust features for the second branch: The maximum energy is achieved at $k_{\text{max}}=k_F=\pi n/2$ and there is a periodicity of $2k_F=\pi n$, where $k_F$ is the Fermi momentum for the noninteracting Fermi gas. These two features are robust against the variations of the dimensionless coupling strength $\gamma_1$ and $\gamma_2$.

\begin{figure}[htp!]
	\includegraphics[scale=0.5]{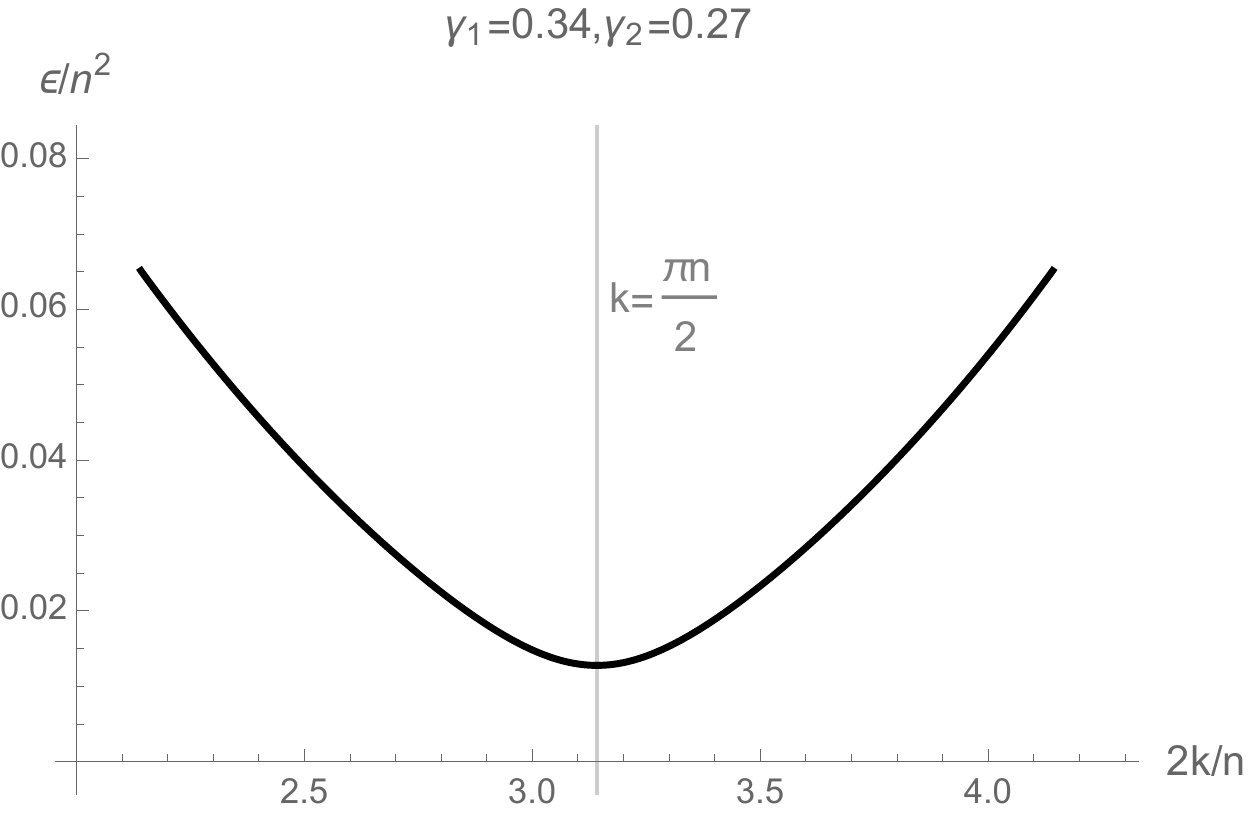}
	\caption{\footnotesize A typical spectrum for the $S=1/2$ excitations. It is similar to that in the Yang-Gaudin model. The minimum energy is fixed at $k_{\text{min}}=k_F=\pi n/2$.}
	\label{fig:dispersion2}
\end{figure}

The spectrum of the $S=1/2$ excitations in the present model is similar to that in the Yang-Gaudin model (see \reffig{fig:dispersion2}). It is gapped and also has a robust extreme: The minimum energy is achieved at $k_{\text{min}}=k_F=\pi n/2$, robust against variations of the dimensionless coupling strength $\gamma_1$ and $\gamma_2$. This is in sharp contrast to the situation in higher dimensions, where the momentum of the minimum energy can be shifted from $k_F$ on the deep BCS side to zero on the deep BEC side \cite{2015qgee.book..179P}. This is also counterintuitive in the sense that there is no conservation law to guarantee this robustness as in the Luttinger theorem, since there is tunneling between atoms and molecules back and forth.

The calculation of low energy excitations for exactly solvable models follows a standard procedure \cite{Samaj_2013,Korepin_1993}. The idea is as follows: The low energy excitations can be excited by various perturbations to the ground state. Since in the thermodynamic limit, the integral equations for the perturbations are linear, we can make a clever choice of linear combinations of the perturbations to give out physical relevant elementary excitations, according to their quantum numbers. Below we carry out this procedure for both $S=0$ excitations and $S=1/2$ excitations.

As in the Lieb-Liniger model, the $S=0$ excitations can be classified into two categories: the hole type and the particle type. For the hole type excitation, a two-string with center momentum $|\Lambda_h|<Q$ is moved to $Q$. This hole type perturbation introduces a shift in the center momentum of the ground state $\Lambda\to \Lambda+\Delta(\Lambda)$. We then define a new function $\omega_h(\Lambda)\equiv \sigma(\Lambda)\Delta(\Lambda)L$, the integral equation for which can be obtained by the usual perturbation theory from \req{eq:discrete}:
\begin{equation}
\begin{split}
\omega_h(\Lambda)-&\frac{1}{2\pi}\int_{-Q}^Qd\Lambda'~\theta'(\Lambda-\Lambda')\omega_h(\Lambda')\\
&=-\frac{1}{2\pi}\left[-\theta(\Lambda-\Lambda_h)+\theta(\Lambda-Q)\right].
\end{split}
\end{equation}
The energy $\epsilon_h$ and momentum $k_h$ of the excitation can be expressed in terms of the function $\omega_h(\Lambda)$:
\begin{equation}
\label{eq:hole}
\begin{split}
\epsilon_h(\Lambda_h)& =\int_{-Q}^Qd\Lambda'~4\Lambda'\omega_h(\Lambda')+2Q^2-2\Lambda_h^2,\\
k_h(\Lambda_h)& =\int_{-Q}^Qd\Lambda'~2\omega_h(\Lambda')+2Q-2\Lambda_h.\\
\end{split}
\end{equation}
For the particle type excitation, a two-string with center momentum $Q$ is moved to $\Lambda_p>Q$. Similarly we define a function $\omega_p(\Lambda)$ and express the energy $\epsilon_p$ and momentum $k_p$ through it:
\begin{equation}
\begin{split}
\omega_p(\Lambda)-&\frac{1}{2\pi}\int_{-Q}^Qd\Lambda'~\theta'(\Lambda-\Lambda')\omega_p(\Lambda')\\
&=-\frac{1}{2\pi}\left[\theta(\Lambda-\Lambda_p)-\theta(\Lambda-Q)\right], \\
\epsilon_p(\Lambda_p)& =\int_{-Q}^Qd\Lambda'~4\Lambda'\omega_p(\Lambda')-2Q^2+2\Lambda_p^2,\\
k_p(\Lambda_p)& =\int_{-Q}^Qd\Lambda'~2\omega_p(\Lambda')-2Q+2\Lambda_p.\\
\end{split}
\end{equation}
These integral equations and dispersion relations can be generally worked out numerically, and a typical result is shown in \reffig{fig:dispersion}. Just as pointed out at the beginning of this subsection, the features of the $S=0$ spectrum are (1) There is a hole branch as well as a particle branch. (2) Both branches are gapless, and at long wavelength they are just phonons with linear dispersion $\epsilon=v_ck$, where $v_c$ is the sound velocity. (3) There are two robust points, the momentum $k_{\text{max}}=k_F$ of the maximum energy and the periodicity $2k_F$.

In fact, the robustness of the periodicity and $k_{\text{max}}$ can be proved from Eqs. (\ref{eq:discrete}) and (\ref{eq:discrete2}) in the thermodynamic limit. Firstly, the periodicity is fixed by the translational invariance: If we shift each $\Lambda$ with the same amount $\pi/L$, then this operation will change the total energy by the amount $NL^{-2}\to 0$, while it will change the total momentum by the amount $(N/2)(2\pi/L)=\pi n$. Secondly, the momentum $k_{\text{max}}$ is fixed by the reflection invariance: If we replace each $\Lambda$ with $\pi/L-\Lambda$, then this operation will also change the total energy by the amount $NL^{-2}\to 0$, such that the spectrum has a reflection symmetry about the total momentum $\pi n/2$.

\begin{figure}
\includegraphics[scale=0.13]{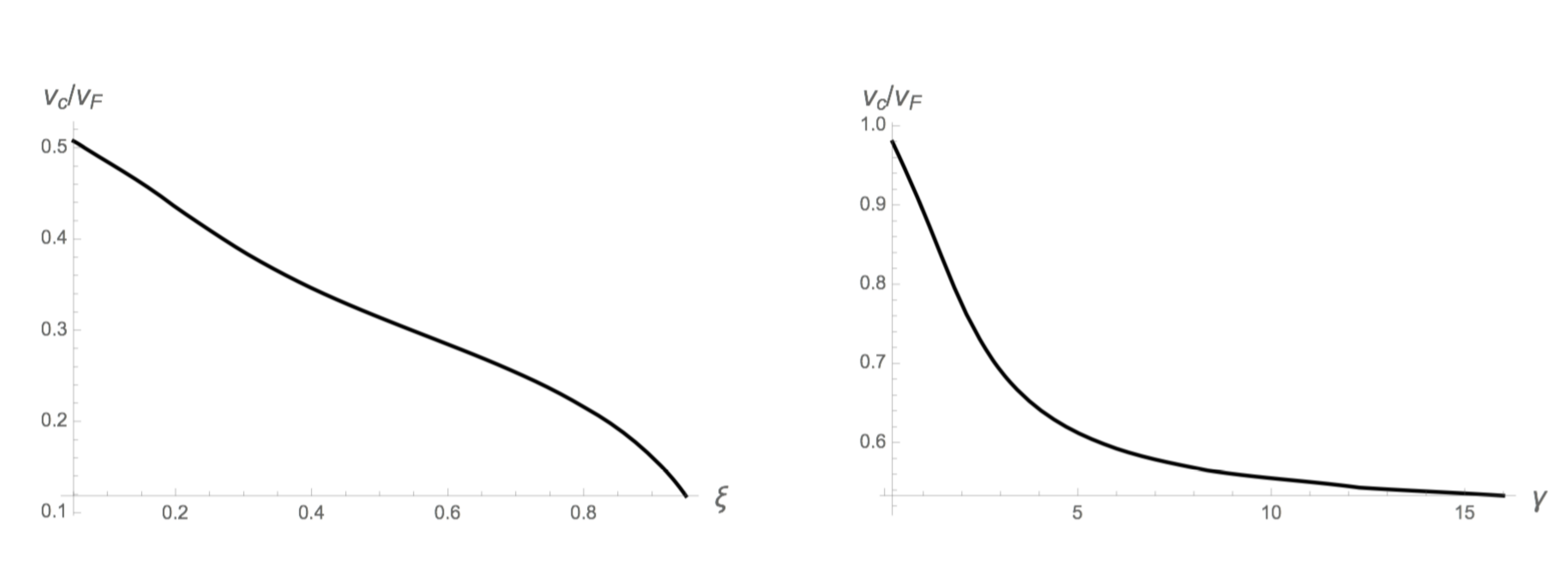}
\caption{\footnotesize Numerical result for sound velocity scaled with $v_F$. Left panel is plotted for the present model with $\gamma_1=5$, where $v_c/v_F$ changes with $\xi=c_2/c_1$. Right panel is plotted for the Yang-Gaudin model by varying dimensionless coupling constant $\gamma=c_F/n$. They are different in the following aspect: The present model has vanishing sound velocity at $\xi\to1$, while the Yang-Gaudin model has a finite lower bound for the sound velocity: $v_c/v_F>0.5$.}
\label{fig:sound}
\end{figure}

After we obtain the spectrum of the $S=0$ excitations, we calculate the sound velocity by linearizing the dispersion $\epsilon(k)$ in the long wavelength limit $k\to 0$. Since in the BCS ($\xi\to 0$) and BEC ($\xi\to 1$) limits the present model reduces to the Yang-Gaudin model and the Lieb-Liniger model respectively (see Eqs. (\ref{eq:YGddd}) and (\ref{eq:LLddd})), the asymptotic behavior of the sound velocity in these two limits can be obtained using the results for the latter two models in the weak coupling limit \cite{PhysRev.130.1616,1975JETP...40..781K,PhysRevLett.93.090408}:
\begin{equation}
\begin{split}
\frac{v_c}{v_F}&\stackrel{\xi\to0}{\approx}1-\frac{\gamma_1}{\pi^2}\xi+\cdots\\
&\stackrel{\xi\to1}{\approx}\frac{\sqrt{\gamma_1(1-\xi)}}{\pi}\left(1-\frac{\sqrt{\gamma_1(1-\xi)}}{2\pi}+\cdots \right),
\end{split}
\end{equation}
where $v_F=\pi n$ is the Fermi velocity for the noninteracting Fermi gas. The sound velocity for general values of $\xi$ is numerically calculated and presented in \reffig{fig:sound}.
We can see that the sound velocity is monotonously decreasing with $\xi$, as the system goes from the BCS ($\xi\to 0$) limit to the BEC ($\xi\to 1$) limit. Also the vanishing of sound velocity in the limit $\xi\to 1$ is consistent with the divergence of compressibility in the same limit. This is in sharp contrast to the Yang-Gaudin model, where the system can never reach the weakly interacting BEC limit.

\begin{figure}[htp!]
	\includegraphics[scale=0.2]{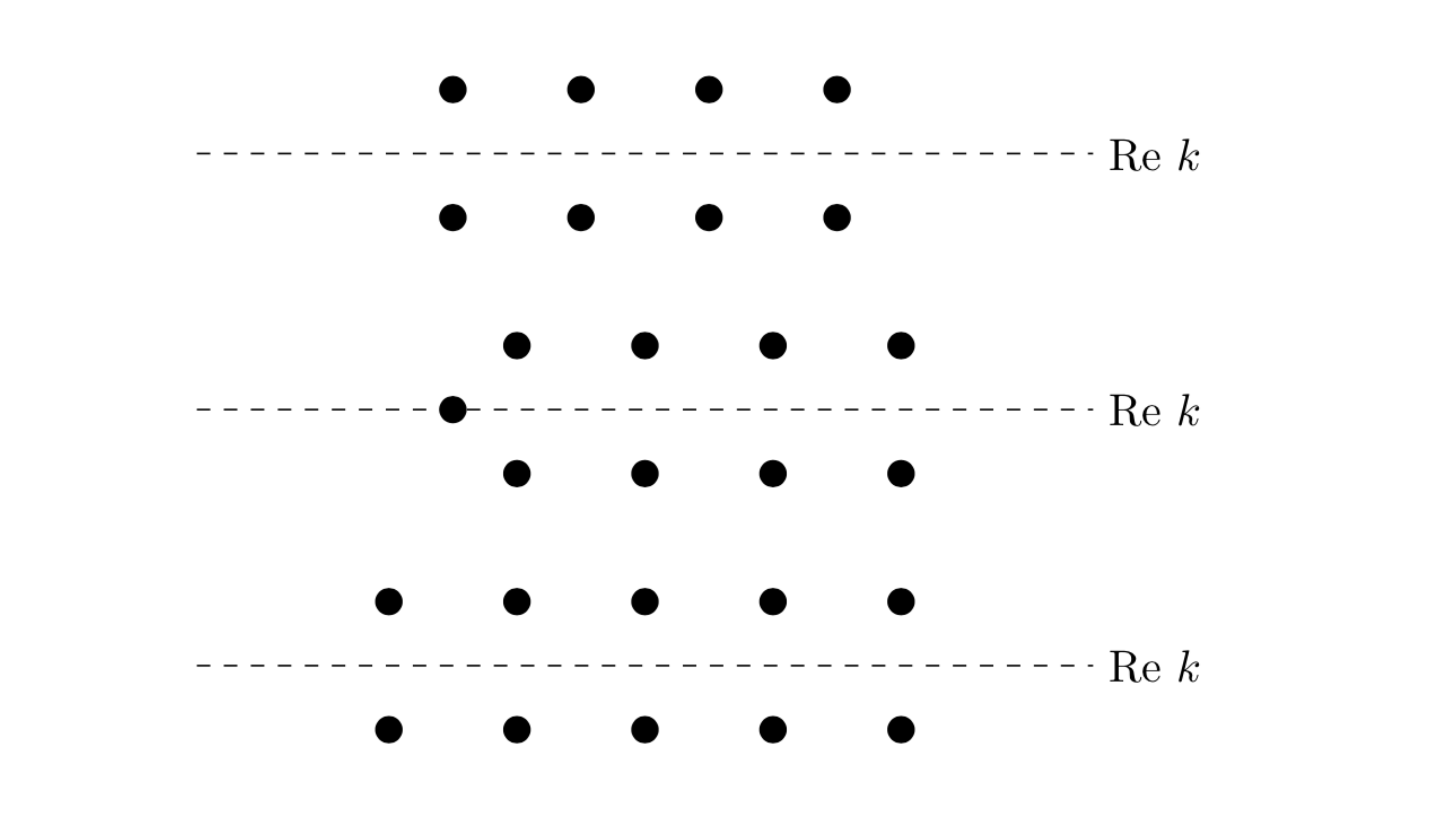}
	\caption{\footnotesize The schematic picture for the distribution of the roots $k$. For two-strings with center momentum $\Lambda_{\alpha}$, the two roots $k_{\alpha,1}$ and $k_{\alpha,2}$ are separated by a distance $c_2$ along the imaginary axis. For unpaired atoms, the roots lie on the real axis. We can see that when a single two-string or a single unpaired atom is added into the system, the $J_{\alpha}$ in \req{eq:discrete} will be shifted by half of unity.}
	\label{fig:rootchange}
\end{figure}

Let us turn to the analysis of the gapped $S=1/2$ excitations. Unlike the $S=0$ excitations, the lowest spin excited state is a triplet state, described by the continuum of two $S=1/2$ excitations and one $S=0$ hole excitation. In this triplet state, we break a two-string with center momentum $|\Lambda_h|<Q$ and add two unpaired atoms with momentum $k_{1,2}$ into the system. The corresponding function $\omega_{\text{tri}}(\Lambda)$ satisfies the following integral equation:
\begin{equation}
\label{eq:trishift}
\begin{split}
&\omega_{\text{tri}}(\Lambda)-\frac{1}{2\pi}\int_{-Q}^Qd\Lambda'~\theta'(\Lambda-\Lambda')\omega_{\text{tri}}(\Lambda')\\
&=\frac{1}{2}-\frac{1}{2\pi}\left[-\theta(\Lambda-\Lambda_h)+\theta_s(\Lambda-k_1)+\theta_s(\Lambda-k_2)\right],
\end{split}
\end{equation}
where the newly defined phase-shift function $\theta_s(\Lambda-k)$ corresponds to the scattering between a molecule and an unpaired atom:
\begin{equation}
\label{eq:BECphase2}
\begin{split}
\theta_s(\Lambda-k)=&-2\arctan\left(\frac{\Lambda-k}{c_2'} \right)+2\arctan\left( \frac{\Lambda-k}{c_1+c_2'} \right)\\
& +2\arctan\left( \frac{\Lambda-k}{c_1-c_2'} \right).
\end{split}
\end{equation}
The extra term $1/2$ on the righthand side of \req{eq:trishift} comes from the fact that the $J_{\alpha}$ in \req{eq:discrete} will change from integer to half-odd integer (or from half-odd integer to integer) if an odd number of two-strings or an odd number of unpaired atoms is added into the system (see Fig. \ref{fig:rootchange}).

The energy $\epsilon_{\text{tri}}$ and momentum $k_{\text{tri}}$ can be once again expressed using the function $\omega_{\text{tri}}(\Lambda)$:
\begin{equation}
\label{eq:triplet2}
\begin{split}
\epsilon_{\text{tri}}&=\int_{-Q}^Qd\Lambda'~4\Lambda'\omega_{\text{tri}}(\Lambda')-2\Lambda_h^2+k^2_1+k^2_2+\frac{c^2_2}{2},\\
k_{\text{tri}}& =\int_{-Q}^Qd\Lambda'~2\omega_{\text{tri}}(\Lambda')-2\Lambda_h+k_1+k_2.
\end{split}
\end{equation}
As we discussed before, Eqs. (\ref{eq:trishift}) and (\ref{eq:triplet2}) are all linear, and this triplet state in fact includes three elementary excitations - one hole type $S=0$ excitation and two $S=1/2$ excitations. By subtracting the $S=0$ component, we are left with the sum of two $S=1/2$ components. There are two ways to define the single $S=1/2$ excitations (we denote them with subscript $s$), both are physically relevant:
\begin{equation}
\label{eq:SpinShift}
\begin{split}
 \omega^{(1)}_s(\Lambda)&-\frac{1}{2\pi}\int_{-Q}^Qd\Lambda'~\theta'(\Lambda-\Lambda')\omega^{(1)}_s(\Lambda')\\
&=\frac{1}{2\pi}\theta(\Lambda-Q)-\frac{1}{2\pi}\theta_s(\Lambda-k),\\
\omega^{(2)}_s(\Lambda)&-\frac{1}{2\pi}\int_{-Q}^Qd\Lambda'~\theta'(\Lambda-\Lambda')\omega^{(2)}_s(\Lambda')\\
&=\frac{1}{2}-\frac{1}{2\pi}\theta_s(\Lambda-k),\\
\end{split}
\end{equation}
The corresponding expressions for energies and momenta are:
\begin{equation}
\label{eq:SpinSpec}
\begin{split}
\epsilon^{(1)}_s &=\int_{-Q}^Qd\Lambda'~4\Lambda'\omega^{(1)}_s(\Lambda')-2Q^2+k^2+\frac{c^2_2}{4}+\mu,\\
k^{(1)}_s& =\int_{-Q}^Qd\Lambda'~2\omega^{(1)}_s(\Lambda')-2Q+k,\\
\epsilon^{(2)}_s &=\int_{-Q}^Qd\Lambda'~4\Lambda'\omega^{(2)}_s(\Lambda')+k^2+\frac{c^2_2}{4}-\mu,\\
k^{(2)}_s& =\int_{-Q}^Qd\Lambda'~2\omega^{(2)}_s(\Lambda')+k,
\end{split}
\end{equation}
For the first definition, we remove a two-string and add an unpaired atom, so the net result corresponds to subtraction of one atom. For the second definition, we only add an unpaired atom (the extra term $1/2$ in the equation for $\omega^{(2)}_s(\Lambda)$ comes from the shift of $J_{\alpha}$ in \req{eq:discrete}, see Fig. \ref{fig:rootchange}), so it corresponds to addition of one atom. Both of them change the number of atoms by one, thus we need to shift their energies by the chemical potential $\mu$, such that the minimum of the two spectra coincides to ensure the particle-hole symmetry of the $S=1/2$ excitations.

\begin{figure}
\includegraphics[scale=0.13]{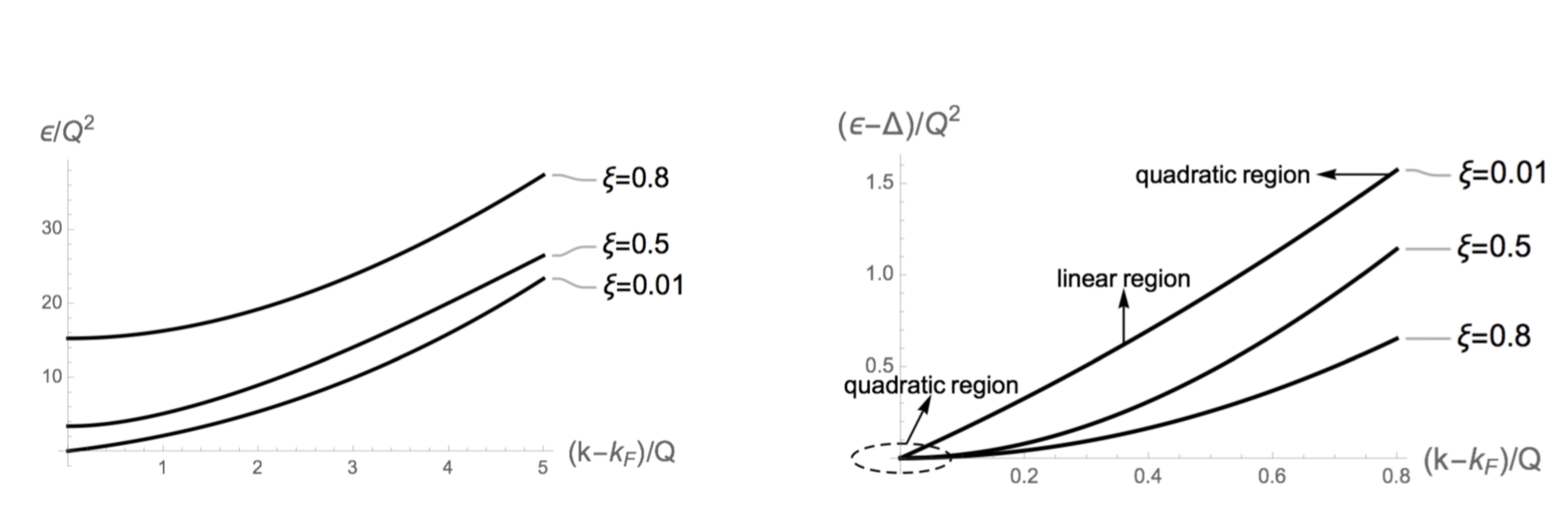}
\caption{\footnotesize Typical spectra for the $S=1/2$ excitations, where the momentum is shifted by $k_F$. Three different choices of $\xi$ are shown.  Right panel is obtained from left panel by offsetting the spin gap. For very small $\xi$, the dispersion curve has three parts: a narrow quadratic region near the minimum, a intermediate linear region and finally a quadratic region at large energy. For $\xi$ close to 1, the dispersion is purely quadratic.}
\label{fig:spinspec}
\end{figure}

\begin{figure}
\includegraphics[scale=0.13]{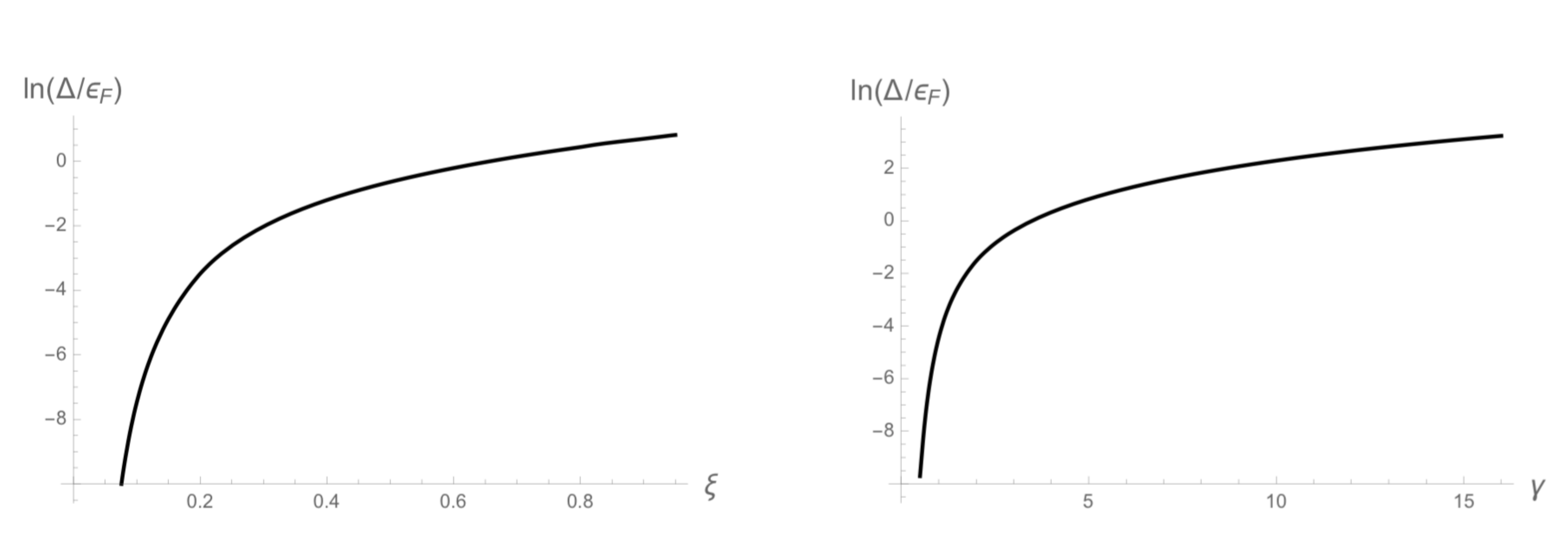}
\caption{\footnotesize Numerical result for spin gap scaled with $\epsilon_F$. Left panel is plotted for the present model with $\gamma_1=5$, and right panel is plotted for the Yang-Gaudin model. They appear practically the same, but with the following difference: the present model terminates at $\xi=1$, while the Yang-Gaudin model will continue the logarithmic behavior with ever growing $\gamma=c_F/n$.}
\label{fig:spingap}
\end{figure}

Solving the above sets of integral equations numerically we will obtain the spectrum for the $S=1/2$ excitations. Typical results are shown in \reffig{fig:spinspec},
where the notable features are: (1) The $S=1/2$ excitations have their lowest energy at momentum $k_{\text{min}}=k_F$, robust against variations of the dimensionless coupling strength $\gamma_1$ and $\gamma_2$. (2) The $S=1/2$ excitations are gapped, where the gap $\Delta(\xi)$ grows with increasing $\xi$ (see \reffig{fig:spingap} for $\Delta(\xi)$ with general values of $\xi$). (3) In the limit $\xi\to 0$, the spectrum is of BCS type, where a small region of quadratic dispersion is followed by an intermediate region of linear dispersion, before the dispersion reaches another quadratic region of large momentum. (4) In the limit $\xi\to 1$, the spectrum is of BEC type, where the dispersion is quadratic all the way.

The robustness of $k_{\text{min}}$ can be verified analytically. For addition of one atom, we can determine the lowest energy configuration by the following condition:
\begin{equation}
\frac{d\epsilon^{(2)}_s}{dk}\Big{|}_{k=k_{\text{min}}}=0.
\end{equation}
This can by solved by variation of the function $\omega^{(2)}_s$ with respect to parameter $k$, and the result is quite simple: $k_{\text{min}}=0$. Now we can calculate the corresponding momentum by substitute $k=0$ into Eqs. (\ref{eq:SpinShift}) and (\ref{eq:SpinSpec}):
\begin{equation}
\label{eq:MimEnergy}
\begin{split}
&  \omega^{(2)}_s(\Lambda)-\frac{1}{2\pi}\int_{-Q}^Qd\Lambda'~\theta'(\Lambda-\Lambda')\omega^{(2)}_s(\Lambda')=\frac{1}{2}-\frac{1}{2\pi}\theta_s(\Lambda),\\
& k^{(2)}_s =\int_{-Q}^Qd\Lambda'~2\omega^{(2)}_s(\Lambda').
\end{split}
\end{equation}
The second term $-\theta_s(\Lambda)/2\pi$ on righthand side of \req{eq:MimEnergy} is odd in $\Lambda$, which has no contribution to the momentum $k_s^{(2)}$, thus we have an alternative expression for $k_s^{(2)}$:
\begin{equation}
\begin{split}
&  \tilde{\omega}_s(\Lambda)-\frac{1}{2\pi}\int_{-Q}^Qd\Lambda'~\theta'(\Lambda-\Lambda')\tilde{\omega}_s(\Lambda')=\frac{1}{2},\\
& k^{(2)}_s =\int_{-Q}^Qd\Lambda'~2\tilde{\omega}_s(\Lambda'),
\end{split}
\end{equation}
This alternative function $\tilde{\omega}_s(\Lambda)$ fulfills the same integral equation as the density of states $\sigma(\Lambda)$, if we make a simple change of the constant inhomogeneous term (see \req{eq:kernel0}). As a result we have:
\begin{equation}
\label{eq:fixedk}
\tilde{\omega}_s(\Lambda)=\frac{\pi}{2}\sigma(\Lambda) \Rightarrow k^{(2)}_s=\pi\int_{-Q}^Qd\Lambda'~\sigma(\Lambda')=\frac{\pi n}{2},
\end{equation}
which means that the minimum of the $S=1/2$ spectrum resides at $k^{(2)}_s=\pi n/2=k_F$ in the case of addition of one atom.

Usually, the fixed momentum $k_F$ appears in the context of Luttinger theorem, which contributes the robustness even in presence of interactions to the conservation of the particle number. In contrast, our result that the minimum of the $S=1/2$ spectrum is fixed at momentum $k_F$ is somewhat surprising, in the sense that the non-conserving nature of the operator $\hat{N}_{\psi}\equiv\int dx ~\hat{\psi}^{\dagger}\hat{\psi}$ (see Eqs. (\ref{eq:Number1}) and (\ref{eq:Number2})) would in principle lead to a changing minimum position in momentum. In fact, the robustness discussed here is due to a special feature of the one dimensional system that the quasiparticle excitation is not stable with respect to soliton formation. Since the full explanation requires a comprehensive semiclassical analysis, which is already beyond the context of the present paper, we will defer it to a later publication \cite{TR2}.

Now we discuss the asymptotic behaviors of the $S=1/2$ excitations in the BCS ($\xi\to 0$) and BEC ($\xi\to 1$) limits. In the BCS limit, the present model reduces to the Yang-Gaudin model (see \req{eq:YGddd}), where the asymptotic behaviors of the spin gap $\Delta$ and the dispersion $\epsilon_s(k)$ near its minimum have already been obtained in the literature in the weak coupling limit \cite{1975JETP...40..781K}:
\begin{equation}
	\frac{\Delta}{\epsilon_F}\approx \frac{8}{\pi}\sqrt{\frac{\gamma_1\xi}{\pi}}e^{-\pi^2/(2\gamma_1\xi)}, ~\epsilon_s(k)\approx \sqrt{\Delta^2+[v_F(k-k_F)]^2}.
\end{equation}
We can see that these results are consistent with the conventional BCS mean field results.

In the BEC limit, the present model reduces to the Lieb-Liniger model with weak dimensionless coupling $\delta\gamma=\gamma_1-\gamma_2\to 0$. In this limit we have $c_2\approx c_1\gg 1$, which makes the second term on the righthand side of \req{eq:SpinShift} for $\omega^{(2)}_s(\Lambda)$ negligible compared with the first term. As a result, we have $\omega^{(2)}(\Lambda)=\pi\sigma(\Lambda)/2$ in the leading approximation, obtained by comparing \req{eq:SpinShift} for $\omega^{(2)}(\Lambda)$ and \req{eq:thermo} for $\sigma(\Lambda)$. Then the asymptotic behaviors of the spin gap $\Delta$ and the dispersion $\epsilon_s(k)$ near its minimum can be obtained from \req{eq:SpinSpec} using \req{eq:LLsigma} for $\sigma(\Lambda)$ in the limit $\xi\to 1$:
\begin{equation}
	\frac{\Delta}{\epsilon_F}\approx \frac{\gamma_1^2}{\pi^2}, ~~~ \epsilon_s\approx \Delta+(k-k_F)^2,
\end{equation}
where the leading term for the spin gap is just the binding energy for the two-strings and the dispersion reduces to free particle form near the minimum of the spectrum. These results are consistent with the usual physical picture of the BEC limit.

Since the presence of the upper bound $Q^*$ for the Fermi momentum has no effect on the low energy excitation spectra, this completes our investigation of the present model in the context of BCS-BEC crossover.

\subsection{Phase Diagram in Presence of External Magnetic Field}
Without an external magnetic field, the ground state of the present model is a Fermi sea of two-strings. By applying external magnetic field above certain threshold depending on the density $n=N/L$, we can polarize the system. Then the ground state will be either fully polarized or mixed with both two-strings and polarized atoms. By varying the magnetic field $H$ and the density $n$, we can explore the phase diagram at zero temperature and observe quantum phase transitions between three different phases: the fully paired ground state (P), the fully polarized ground state (FP) and the partially polarized ground state (PP).  This kind of analysis has already been done for the Yang-Gaudin model in the literature \cite{PhysRevLett.98.070402,PhysRevLett.98.070403,PhysRevB.76.085120,PhysRevA.84.023616}. It is pointed out that the PP phase is gapless, and the power-law decay of the pair correlation $\langle \psi^{\dagger}_{\uparrow}(x)\psi^{\dagger}_{\downarrow}(x)\psi_{\uparrow}(0)\psi_{\downarrow}(0) \rangle\propto \cos\left(k_{\text{FFLO}}|x|\right)/|x|^{\alpha}$ is accompanied by a spatial oscillation. The wave vector of this oscillation was numerically found to depend on the mismatch of the Fermi points $k_{\text{FFLO}}\simeq\pi(n_{\uparrow}-n_{\downarrow})$. Thus the PP phase serves as the one dimensional analog of the Fulde-Ferrell-Larkin-Ovchinnikov (FFLO) state, and it provides an ideal place to find and explore the superfluid phase with inhomogeneity.

In this section, we calculate the zero temperature phase diagram of the present model. For technical convenience, we start from the grand canonical ensemble, where the chemical potential $\mu$ is introduced as an auxiliary parameter to establish the phase boundaries in the $H-n$ space. Also for definiteness, we choose the case $H\geqslant 0$.

We introduce two density of states, one for the unpaired atoms, which is denoted as $\rho(k)$ and one for the molecules, which is denoted as $\sigma(\Lambda)$. Then we have
\begin{equation}
\label{eq:norm2}
\begin{split}
	n_{\uparrow}+n_{\downarrow}=&2\int_{-Q}^Qd\Lambda~\sigma(\Lambda)+\int_{-q}^qdk~\rho(k),\\
	n_{\uparrow}-n_{\downarrow}=&\int_{-q}^qdk~\rho(k),
\end{split}
\end{equation}
where $m_{\psi}(n_{\uparrow}+n_{\downarrow})$ is the total mass density, $(n_{\uparrow}-n_{\downarrow})$ is the total spin density, $q$ is the Fermi momentum of the unpaired atoms and $Q$ is the Fermi momentum of the molecules. Following the same procedure as that in deriving \req{eq:kernel0}, we obtain the coupled equations for the two density of statess:
\begin{equation}
\label{eq:norm3}
	\begin{split}
		& \rho(k)=\frac{1}{2\pi}+\frac{1}{2\pi}\int_{-q}^qdk'~\theta'_{ss}(k-k')\rho(k')\\
		& ~~~~~~~~~~~~~+\frac{1}{2\pi}\int_{-Q}^Qd\Lambda~\theta'_s(k-\Lambda)\sigma(\Lambda),\\
		& \sigma(\Lambda)=\frac{1}{\pi}+\frac{1}{2\pi}\int_{-q}^qdk~\theta'_s(\Lambda-k)\rho(k)\\
		&~~~~~~~~~~~~~+\frac{1}{2\pi}\int_{-Q}^Qd\Lambda'~\theta'(\Lambda-\Lambda')\sigma(\Lambda'),
	\end{split}
\end{equation}
where the phase-shift functions $\theta(\Lambda-\Lambda')$ and $\theta_s(\Lambda-k)$ are defined in Eqs. (\ref{eq:smooth}) and (\ref{eq:BECphase2}) respectively, and the new phase-shift function
\begin{equation}
	\theta_{ss}(k-k')=2\arctan\left(\frac{k-k'}{c_1} \right)
\end{equation}
corresponds to the scattering between two unpaired atoms with the same spin direction. The ground state energy density $\epsilon$ of the system is then
\begin{equation}
\begin{split}
	\epsilon=\frac{E}{L}=&\int_{-Q}^Qd\Lambda \left(2\Lambda^2-\frac{c^2_2}{2} \right)\sigma(\Lambda)+\int_{-q}^qdk~k^2\rho(k).
\end{split}
\end{equation}

Performing variation of $\epsilon$ with respect to $\sigma(\Lambda)$ and $\rho(k)$ under the constraint in \req{eq:norm2} and making use of \req{eq:norm3}, we obtain
\begin{equation}
\begin{split}
\epsilon^u(k)=& k^2-\mu-h+\frac{1}{2\pi}\int_{-q}^qdk'~\theta'_{ss}(k-k')\epsilon^u(k')\\
&+\frac{1}{2\pi}\int_{-Q}^Qd\Lambda~\theta'_s(k-\Lambda)\epsilon^b(\Lambda), \\
\epsilon^b(\Lambda)= & 2\left( \Lambda^2-\mu-\frac{c^2_2}{4}\right)+\frac{1}{2\pi}\int_{-q}^qdk~\theta'_s(\Lambda-k)\epsilon^u(k),\\
& +\frac{1}{2\pi}\int_{-Q}^Qd\Lambda'~\theta'(\Lambda-\Lambda')\epsilon^b(\Lambda'),
\end{split}
\end{equation}
where the chemical potential $\mu$ and the reduced magnetic field $h=H/2$ are the two Lagrange multipliers. The two introduced functions $\epsilon^u(k)$ and $\epsilon^b(\Lambda)$ are referred to as the dressed energy functions for the unpaired atoms and molecules respectively \cite{Takahashi_1999,PhysRevB.76.085120}. The dressed energy function is introduced in the grand canonical ensemble such that it gives out negative value when the momentum is within the Fermi sea and positive value when the momentum is outside the Fermi sea. Equivalently, we have the condition that
\begin{equation}
\epsilon^u(\pm q)=0, ~~~ \epsilon^b(\pm Q)=0.
\end{equation}

\begin{figure}
\includegraphics[scale=0.18]{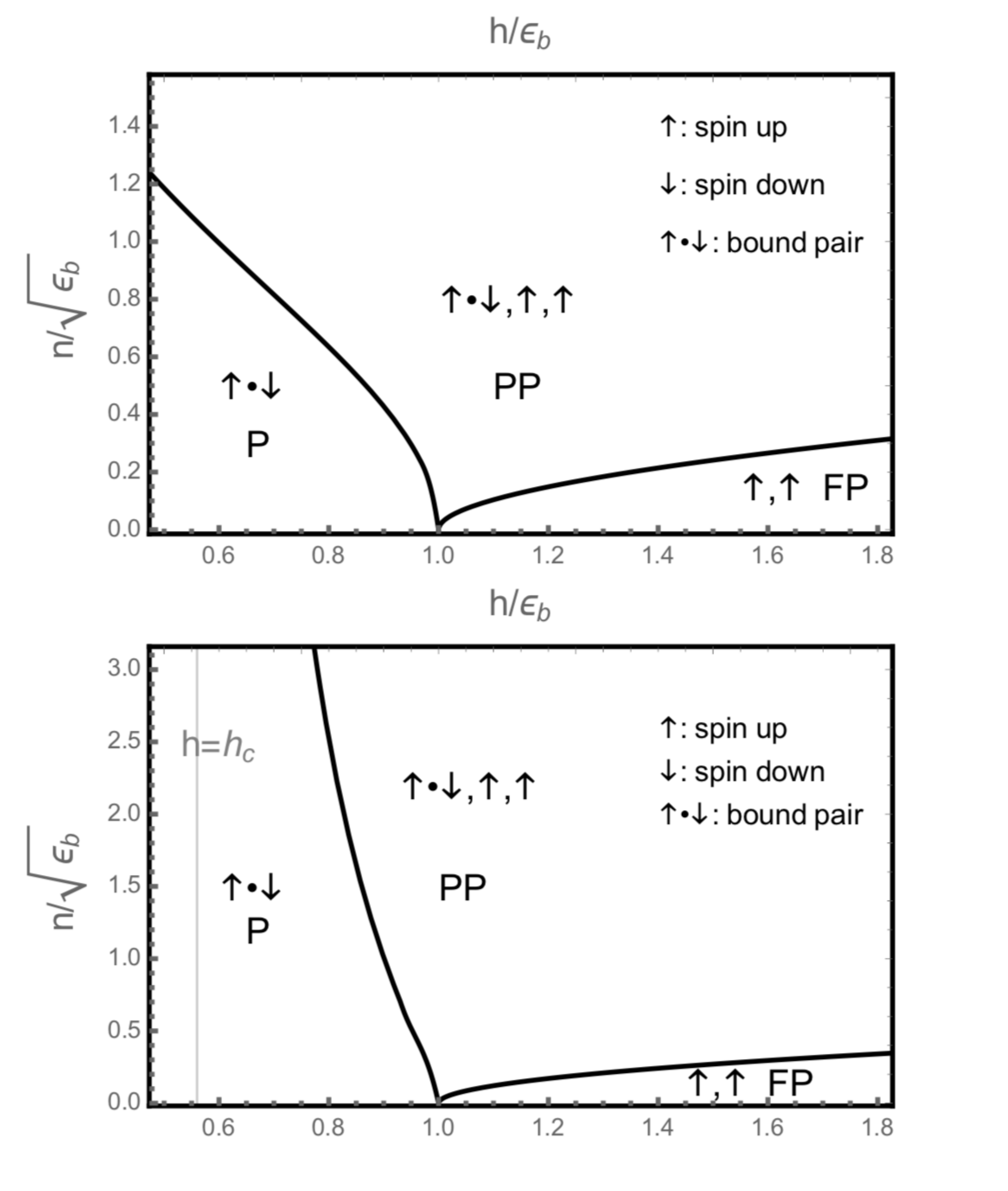}
\caption{\footnotesize Phase diagram in the $h-n$ space at zero temperature, where $n$ and $h$ are scaled by $\epsilon_b=c^2_2/4$. Upper panel is plotted for $\xi=0.1$, and lower panel is plotted for $\xi=0.8$. The case with $\xi=0.1$ is essentially the same as that in the Yang-Gaudin model, while in the case with $\xi=0.8$, the mixed phase region (PP) is reduced. The phase boundary P-PP actually has an asymptote corresponding to the critical magnetic field $h=h_c$.}
\label{fig:nhphase}
\end{figure}

In terms of the dressed energies, the zero temperature phase diagram can be calculated as follows. The boundary from fully polarized to partially polarized ground state is determined by the condition
\begin{equation}
\epsilon^u(\pm q)=0, ~~~ \epsilon^b(0)\leqslant 0.
\end{equation}
Then the phase boundary FP-PP can be obtained as the solution $n=n_1(h)$ to the coupled integral equations:
\begin{equation}
	\begin{split}
		& \epsilon^u(k)=k^2-\mu-h+\frac{1}{2\pi}\int_{-q}^qdk'~\theta'_{ss}(k-k')\epsilon^u(k'), \\
		& \mu=-\frac{c^2_2}{4}+\frac{1}{4\pi}\int_{-q}^qdk~\theta'_s(-k)\epsilon^u(k),\\
		& \rho(k)=\frac{1}{2\pi}+\frac{1}{2\pi}\int_{-q}^qdk'~\theta'_{ss}(k-k')\rho(k'), \\
		& n=\int_{-q}^qdk~\rho(k).
	\end{split}
\end{equation}
The boundary from fully paired to partially polarized ground state is determined by the condition
\begin{equation}
\epsilon^u(0)\leqslant 0, ~~~ \epsilon^b(\pm Q)=0.
\end{equation}
Then the phase boundary P-PP can be obtained as the solution $n=n_2(h)$ to the coupled integral equations:
\begin{equation}
	\begin{split}
		& \epsilon^b(\Lambda)=2\left( \Lambda^2-\mu-\frac{c^2_2}{4}\right)+\frac{1}{2\pi}\int_{-Q}^Qd\Lambda'~\theta'(\Lambda-\Lambda')\epsilon^b(\Lambda'), \\
		& \mu=-h+\frac{1}{2\pi}\int_{-Q}^Qd\Lambda~\theta'_s(-\Lambda)\epsilon^b(\Lambda), \\
		& \sigma(\Lambda)=\frac{1}{\pi}+\frac{1}{2\pi}\int_{-Q}^Qd\Lambda'~\theta'(\Lambda-\Lambda')\sigma(\Lambda'), \\
		& n=2\int_{-Q}^Qd\Lambda~\sigma(\Lambda).
	\end{split}
\end{equation}
The functions $n_1(h)$ and $n_2(h)$ cannot be expressed in closed forms, they can only be obtained by directly dealing with the corresponding coupled integral equations, generally through numerical calculations. Typical phase diagrams are presented in \reffig{fig:nhphase}, where $n$ and $h$ are scaled by $\sqrt{\epsilon_b}$ and $\epsilon_b$ respectively, with $\epsilon_b=c^2_2/4$ being the binding energy per atom. Comparison is made between $\xi\ll 1$ and $\xi\sim 1$ - The phase diagram at small $\xi$ is essentially the same as that in the Yang-Gaudin model. When $\xi$ goes near 1, the phase diagram develops a new feature: there arises a critical strength $h_c$ for the magnetic field, below which the ground state is always a Fermi sea of two-strings and cannot be polarized.

The critical magnetic field can be shown to come about due to the presence of upper bound $Q^*$ on the Fermi momentum of the system. For fixed $c_1$ and $c_2$, when we increase the chemical potential or the mass density, we will finally get close to the upper bound $Q^*$. In the range $0.4<\xi<1$, the density of states $\sigma(x)$ is then dominated by square root singular term $\sigma^0(x)$ in \req{eq:zeromode}. We then use it together with \req{eq:ENperL} to calculate the leading order contribution to the energy density $\epsilon\equiv E/L$ when $Q$ approaches $Q^*$ from below:
\begin{equation}
	\begin{split}
		\epsilon&=\frac{E}{L}\approx {Q^*}^3\int_{-1}^1dx\left(2x^2-\frac{\xi^2{\lambda^*}^2}{2} \right)\sigma^0(x)\\
		&=\frac{\pi{Q^*}^3F(Q^*)(1-\xi^2{\lambda^*}^2)}{4(Q^*-Q)}, \\
		\frac{n}{2}&=\frac{N}{2L}\approx Q^*\int_{-1}^1dx~\sigma^0(x)=\frac{\pi Q^*F(Q^*)}{2(Q^*-Q)},
	\end{split}
\end{equation}
where $\xi=c_2/c_1$, $\lambda^*=c_1/Q^*$, and the relation between $\lambda^*$ and $\xi$ can be read off from \req{eq:fit} or \reffig{fig:critical}.
This shows that the leading order contribution to the energy density is linear in $n=N/L$:
\begin{equation}
\label{eq:noninteracting}
\epsilon(n)\approx B(\xi)n, ~~~ B(\xi)=\frac{1/{\lambda^*}^2-\xi^2}{4}c^2_1.
\end{equation}
For fixed number of particles $n=n_{\uparrow}+n_{\downarrow}$ with small varying polarization $\delta n=n_{\uparrow}-n_{\downarrow}$, the energy of the system can be expressed as
\begin{equation}
\epsilon(\delta n)=B(\xi)n-(h+B(\xi))\delta n.
\end{equation}
When $\xi$ is small, the coefficient $B(\xi)$ is positive, which means that an infinitesimal magnetic field will polarize the system as long as the mass density of the system is large enough, and there is no critical magnetic field $h_c$. When $\xi$ goes to 1, $\lambda^*$ tends to diverge, then we have $B(\xi)<0$, which means that we need a finite strength of magnetic field $h_c=-B(\xi)$ to polarize the system. The critical $\xi^*$ is then determined by
\begin{equation}
B(\xi^*)=0 \Rightarrow \lambda^*(\xi^*)\xi^*=1 \Rightarrow \xi^*=0.61.
\end{equation}
The value of $\xi^*$ falls in the range $0.4<\xi<1$, so the usage of the square root singular form in \req{eq:zeromode} for the density of states when $Q$ approaches $Q^*$ from below is justified.

As a result, we have $h_c=0$ for $\xi<\xi^*$ and $h_c=-B(\xi)$ for $\xi>\xi^*$. If we approach the critical value $\xi^*$ from above, the critical magnetic field will display the following critical behavior:
\begin{equation}
h_c \sim (\xi-\xi^*)^{\alpha_h} ~~~ \text{for}~~~ \xi=\xi^*+0,
\end{equation}
where the critical exponent $\alpha_h$ can be calculated from the functional form of $B(\xi)$ with the result $\alpha_h=1$. This result comes from the fact that the system can be viewed as a collection of noninteracting particles in the leading approximation according to \req{eq:noninteracting}. Since \req{eq:noninteracting} is obtained by keeping only the singular part $\sigma^0(x)$ from the density of states $\sigma(x)=\sigma_{\text{reg}}(x)+\sigma^0(x)$, the interaction effect comes from the regular part $\sigma_{\text{reg}}(x)$, which produces a higher order correction to the result $\alpha_h=1$.

\section{Bright Solitons with $c_1<c_2$}
\label{sec:string}
\begin{figure}[htp!]
	\includegraphics[scale=0.2]{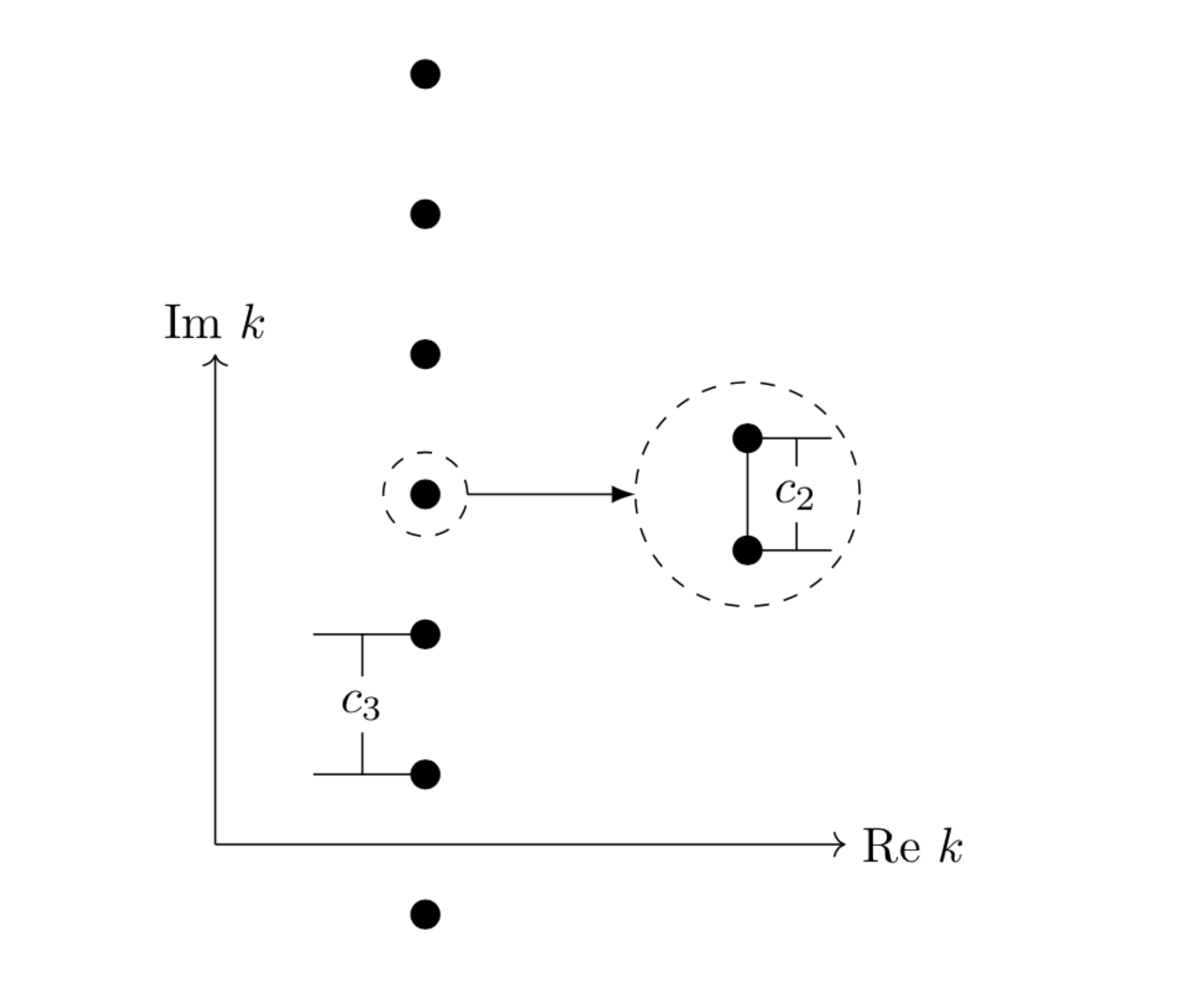}
	\caption{\footnotesize The embedded string solution. The two-string (the one in the enlarged circle) with inter-root separation $c_2$ is embedded in the $M$-string with inter-root separation $c_3=c_2-c_1$.}
	\label{fig:embed}
\end{figure}

In the previous section, we have touched the issue that a collapsing instability appears when we tune $c_2$ close to $c_1$, see Figs. (\ref{fig:critical}), (\ref{fig:compress}) and (\ref{fig:sound}). In this section, we focus on the regime $0<c_1<c_2$, the instability discussed in the previous section implies we would have a collapsing solution in this regime for fermionic atoms (see \req{eq:collapse}). This counterintuitive result is due to the fact that the fermionic atoms are tightly bound into bosonic molecules with residual attraction before collapsing. In this section, we confirm this claim. Firstly we still make the two-string ansatz like that in \req{eq:twostring}
\begin{equation}
k_{\alpha,1}=\Lambda_{\alpha}+iv, ~~~ k_{\alpha,2}=\Lambda_{\alpha}-iv, ~~~ v>0,
\end{equation}
where $\alpha=1,2,\cdots,M=N/2$. But this time we leave the reality of the center momentum $\Lambda_{\alpha}$ for the moment. The Bethe ansatz equations in \req{eq:Bethe} then implies
\begin{equation}
\exp(-2vL)\sim \left( \frac{v-c'_2}{v+c'_2} \right)^2.
\end{equation}
For a macroscopic system where $L\to\infty$, this fixes the value $v=c'_2$, and \req{eq:Bethe2} still follows. This time we have $c_3=c_2-c_1>0$, and we have the following Bethe equations for $\Lambda_{\alpha}$:
\begin{equation}
\label{eq:eqstring}
\exp(i2\Lambda_{\alpha}L)\sim \prod_{\beta=1}^M\left(\frac{\Lambda_{\alpha}-\Lambda_{\beta}-ic_3}{\Lambda_{\alpha}-\Lambda_{\beta}+ic_3}\right),
\end{equation}
where we have omitted other factors which have no effect on the subsequent derivations \footnote{The second and the third terms in \req{eq:Bethe2} cannot produce physical poles; The first term in \req{eq:Bethe2} can produce physical poles, but it won't lead to a valid string solution, because it will lead to repeating values in original momentum $k$, then the exclusion principle in one dimension tells us that the wave function would vanish.}. Now \req{eq:eqstring} has the same form as that appearing in the attractive Lieb-Liniger model, whose ground state is a single string solution encompassing all particles \cite{PhysRev.130.1605,Samaj_2013}. Subsequently, unlike the uniform regime, we now have a single $M$-string solution for center momentum $\Lambda_{\alpha}$:
\begin{equation}
\Lambda_{\alpha}=u+ic'_3(M+1-2\alpha), ~~~ c_3'=c_3/2,
\end{equation}
where $u$ is a real number and $\alpha=1,2,\cdots,M=N/2$. We now have an embedded string solution of the following structure
\begin{equation}
\label{eq:embedd}
\begin{split}
& k_{\alpha,1}=\Lambda_{\alpha}+ic_2', ~~~ k_{\alpha,2}=\Lambda_{\alpha}-ic_2',\\
& \Lambda_{\alpha}=u+ic'_3(M+1-2\alpha),
\end{split}
\end{equation}
where the label $\alpha$ runs from $1$ to $M=N/2$. The physical picture of this embedded string solution is a bound state encompassing all particles: Firstly, atoms with opposite spins are bound into molecules, then the molecules are bound together as a single bright soliton due to the residual attraction (see \reffig{fig:embed}).

The above physical picture can be better understood by writing down the corresponding wave functions directly. This can be done through the nested coordinate Bethe ansatz:
\begin{equation}
\label{eq:nestBA1}
\small
\begin{split}
& \Psi_{X}(\sigma_1,x_1;\cdots;\sigma_N,x_N)=\sum_{P\in S_N}[X,P]\exp\left( i\sum_{j=1}^Nk_{P_j}x_{X_j} \right), \\
& [X,P]=\text{sgn}(X)\text{sgn}(P)A_{\sigma_{X_1}\sigma_{X_2}\cdots\sigma_{X_N}}(k_{P_1},k_{P_2},\cdots,k_{P_N}),
\end{split}
\end{equation}
where $\sigma_i=\uparrow,\downarrow$ denotes the spin directions, $X$ denotes the ordering sector with $x_{X_1}<x_{X_2}<\cdots<x_{X_N}$ and $P$ denotes the permutation among the wave numbers. The sign function equals 1 for even permutations and -1 for odd permutations. For $M$ down-spins sitting at integer positions $1\leqslant y_1<y_2<\cdots<y_M\leqslant N$, we denote the function $A_{\sigma_{X_1}\cdots\sigma_{X_N}}(k_{P_1},\cdots,k_{P_N})$ as $A_P(y_1,y_2,\cdots,y_M)$.  This function is obtained by generalizing the result for the Yang-Gaudin model \cite{GAUDIN196755,PhysRevLett.19.1312} from a single coupling constant $c_F$ to two coupling constants $c_1$ and $c_2$:
\begin{equation}
\label{eq:nestBA2}
\begin{split}
& A_P(y_1,y_2,\cdots,y_M)=\sum_{R\in S_M}G_1(P)G_2(R)\prod_{i=1}^MF_P(y_i,\Lambda_{R_i}), \\
& G_1(P)=\prod_{j<l}(k_{P_j}-k_{P_l}+ic_1),\\
& G_2(R)=\text{sgn}(R)\prod_{j<l}(\Lambda_{R_j}-\Lambda_{R_l}+ic_2), \\
& F_P(y,\Lambda)=\prod_{j=1}^{y-1}(k_{P_j}-\Lambda-ic'_2)\prod_{l=y+1}^N(k_{P_l}-\Lambda+ic'_2).
\end{split}
\end{equation}
For a concrete illustration, we substitute \req{eq:embedd} into Eqs. (\ref{eq:nestBA1}) and (\ref{eq:nestBA2}) in the case with $N=4$, then in the basic sector $I:x_1<x_2<x_3<x_4$, we have:
\begin{equation}
\small
\begin{split}
& \Psi_I(\uparrow,x_1;\downarrow,x_2;\uparrow,x_3;\downarrow,x_4)\sim e^{-c'_2(x_2+x_4-x_1-x_3)-c'_3(x_3+x_4-x_1-x_2)}, \\
& \Psi_I(\uparrow,x_1;\uparrow,x_2;\downarrow,x_3;\downarrow,x_4) =0,\\
& \Psi_I(\uparrow,x_1;\downarrow,x_2;\downarrow,x_3;\uparrow,x_4)\sim e^{-c'_2(x_2+x_4-x_1-x_3)-c'_3(x_3+x_4-x_1-x_2)},
\end{split}
\end{equation}
where we have set $u=0$ for simplicity. The expression for other ordering sectors then follows from symmetry of the system. Through the explicit wave function, the physical picture of the embedded string solution is transparent, where the exponential decay on the length scale of $1/c'_2$ represents the molecule structure and the exponential decay on the length scale of $1/c'_3$ binds all the $M$ molecules together.

\section{Conclusion}
In this paper, we introduced models of two-component bosons and fermions with tunable inter-species interactions in one dimension, which is subject to exact solutions by Bethe ansatz upon fine-tuning. The tunable interactions are realized by Feshbach resonances of two antiparallel pseudospins. The $N$ atoms in this model can be imagined to live on hyperplanes corresponding to different ordering sectors $X:x_{X_1}<x_{X_2}, <\cdots<x_{X_N}$. Without reaction that converts atoms and molecules back and forth, we only need to require the continuity of wave function when hyperplanes intersect. With Feshbach resonances, the molecules can be viewed as living on the intersections of the hyperplanes, which play the role of the boundary conditions. The resulting Bethe ansatz equations admit two types of ground state solutions, depending on the relation between the two coupling constants $c_1,c_2$. In the regime $c_1>c_2$, the ground state is a Fermi sea of two-strings, where the Fermi wave vector $Q$ is under constraint: there is a limiting $Q^*$ which $Q$ cannot exceed, and near $Q^*$ the distribution of the center momentum presents a square root singularity. As $c_2$ approach $c_1$ from below we come close to a diverging compressibility, which leads us into the other regime $c_2>c_1$ with a single $N$ particle bound state for the ground state. In the Bethe approach, this bound state reveals itself as an embedded string solution. Our model is experimentally accessible by using two hyperfine states of $^{87}Rb$ quantum gases with tunable couplings via Feshbach resonance. Furthermore, the uniform regime $c_1>c_2$ provides a new scenario for investigating the physics of BCS-BEC crossover, where the system is governed by a single Hamiltonian and the behavior of the spin excitations is accessible along the whole range of the crossover. Also, we have explored the zero temperature phase diagram in presence of external magnetic field, where a critical magnetic field below which the ground state cannot be polarized is caused by the presence of the upper bound $Q^*$ for the Fermi momentum.

\section{Acknowledgment}
We would like to thank Victor Gurarie, Yang-Zhi Chou and Natan Andrei for helpful discussions and valuable comments on the paper. This work is supported by Simons Foundation.

\appendix
\numberwithin{equation}{section}

\bibliography{TwoComponent}

\end{document}